\documentclass[aps,prx,twocolumn,superscriptaddress]{revtex4-1}
\usepackage{amsmath}
\usepackage{amssymb}
\usepackage{color}
\usepackage{graphicx}
\usepackage{epstopdf}

\DeclareMathOperator{\arcosh}{arcosh}
\DeclareMathOperator{\arsinh}{arsinh}
\DeclareMathOperator{\artanh}{artanh}
\DeclareMathOperator{\arcoth}{arcoth}

\begin{document}

\title{Surface potentials of conductors in electrolyte solutions  }

\author{Olga I. Vinogradova}
\email[Corresponding author: ]{oivinograd@yahoo.com}
\affiliation{Frumkin Institute of Physical Chemistry and Electrochemistry, Russian Academy of Sciences, 31 Leninsky Prospect, 119071 Moscow, Russia}

\author{Elena F. Silkina}
\affiliation{Frumkin Institute of Physical Chemistry and Electrochemistry, Russian Academy of Sciences, 31 Leninsky Prospect, 119071 Moscow, Russia}

\author{Evgeny S. Asmolov}
\affiliation{Frumkin Institute of Physical Chemistry and Electrochemistry, Russian Academy of Sciences, 31 Leninsky Prospect, 119071 Moscow, Russia}
\affiliation{Lomonosov Moscow State
University, 119991 Moscow, Russia}

\begin{abstract}
When we  place conducting bodies in electrolyte solutions, their surface potential $\Phi_s$ appears to be much smaller in magnitude than the intrinsic one $\Phi_0$  and  normally does not obey the classical electrostatic boundary condition of a constant surface potential  expected for conductors. In this paper, we demonstrate that  an explanation of these observations can be obtained
by postulating  that diffuse ions condense at the ``wall'' due to a reduced permittivity of a solvent.
 For small values of $\Phi_0$ the surface potential responds linearly. On  increasing $\Phi_0$ further $\Phi_s$ augments nonlinearly and then saturates to a constant value. Analytical approximations for $\Phi_s$ derived for these three distinct modes  show that it always adjusts to salt concentration, which is equivalent to a violation of the constant potential condition. The latter would be  appropriate for highly dilute solutions, but only if  $\Phi_0$ is small. Surprisingly, when the plateau with high $\Phi_s$  is reached, the  conductor surface switches to a constant charge density condition normally expected for insulators. Our results are directly relevant for conducting electrodes, mercury drops, colloidal metallic particles and more.
\end{abstract}

\maketitle

\section{Introduction}

The concept of the electric double layer in an electrolyte solution adjacent to a charged bodies is one of the central in colloid science and electrochemistry, since it is an origin of numerous phenomena. The classical model assumes that the electric double layer consists of an inner Stern layer of a thickness $\delta$ below a few molecular sizes and a so-called  outer electrostatic
diffuse layer (EDL) that extends to distances of the order of the Debye
length $\lambda$ of a bulk electrolyte solution.   The potential  $\Phi_s$ at the boundary
between the Stern and diffuse layers is termed a surface potential~\cite{stern.o:1924}.
The latter is generally much smaller than the  intrinsic (applied) potential $\Phi_0$ of the conducting body~\cite{connor.jn:2001} we refer below as a ``wall'' (or electrode) potential. The Stern layer plays an important role in capacitive
processes (electrowetting, desalination)~\cite{mott.nf:1961,conway.be:1999} and in Faradaic electrode processes, such as the
kinetics of anode and cathode reactions in fuel cells~\cite{biesheuvel.pm:2021}. Contrary to that, an outer part of the double layer determines the coagulation of colloidal suspensions and electrokinetic phenomena, thus to quantify them the surface potential only is required~\cite{derjaguin.bv:1941,vervey:1948,israelachvili.jn:2011,anderson.jl:1989,nizkaya.tv:2022,vinogradova.oi:2024}.
Over the past century the notion of an electric double layer has become widely accepted, but that does not mean that interpretation of its ingredients is fully established.

It is now well accepted that the mean-field Poisson-Boltzmann equation, which assumes point-like ions in thermodynamic equilibrium and neglects correlations, describes the spatial distribution of electrostatic potential in the EDL~\cite{andelman.d:2006book,joly.l:PB}, and some direct experimental (qualitative~\cite{bedzyk.mj:1990} and quantitative~\cite{connor.jn:2001}) validations have been provided in the last decades. What, however, remains unclear is how $\Phi_s$ is established and why numerous observations in electrolytes are in conflict with expected for conductors a  constant surface potential (Dirichlet) boundary condition. It is natural to suggest that the Stern layer, if any, does induce $\Phi_s$ self-consistently and by this reason should be incorporated to the model even of those phenomena that are formally controlled by an outer EDL.
For the Stern layer, a situation is much less clear both on the theoretical and on the experimental sides. Since long time it was axiomatic in colloid science that in this inner layer the dielectric constant  is identical to its bulk value, but the use of the Poisson-Boltzmann equation is poorly justified~\cite{hartkamp.r:2018}.
To explain the voltage drop inside the Stern layer a specific (i.e. non-Coulombic) adsorption of potential-determining ions is traditionally involved~\cite{lyklema:2010}. Such ions are treated as strongly chemisorbed and, consequently, immobile, in contrast to ions in the bulk and outer diffuse layer. Another point of view is that the Stern layer is charge-free and its existence simply reflects a closest-approach distance for (the centers of) the real ions to come near the ``wall''~\cite{biesheuvel.pm:2021}. These assumptions have no direct experimental evidences because it is still impossible to probe the ionic distribution so close to the ``wall''. Thus one can speculate that for many substrates the Stern layer properties may be remote from  the standard colloid models.

Several apparently unrelated developments suggest this. The first of these has to do with the permittivity of solvent near the ``wall''. To interpret the double layer capacitance, electrochemists have long speculated that the permittivity in the Stern layer is steeply decreased owing to the orientation of adsorbed solvent dipoles~\cite{conway.be:1951}.
This picture is consistent with more recent macroscopic theories~\cite{borisevich.sv:1999}, molecular models treated by methods of integral equations~\cite{guidelli.r:2000} and with computer simulations~\cite{hartkamp.r:2018,schlaich.a:2016,becker.m:2023}, where the same conclusions have been drawn.
Recent direct experiments have demonstrated that permittivity of pure water indeed abruptly drops at subnanometer distances from the ``wall''~\cite{fumagalli.l:2018}.
The second development has to do with the computer simulations of specific adsorption of potential-determining ions~\cite{maduar.sr:2015,grosjean.b:2019}. These computer experiments have revealed the existence of an adsorbed layer of ions being in thermodynamic equilibrium with the bulk solution. Although the thickness of such a weakly physisorbed layer is only about one to two ion diameters, the (lateral) mobility of ions remains large~\cite{mouterde.t:2019}.  Such a mobility of adsorbed ions significantly  impacts electrokinetic transport in micro- and nanofluidic channels~\cite{maduar.sr:2015,mouterde.t:2018,silkina.ef:2019,vinogradova.oi:2022,mangaud.e:2022,vinogradova.oi:2023}. We also note that the classical assumption of immobile specifically chemisorbed ions is  not physically realistic at a mercury electrode, where they are certainly mobile~\cite{conway.be:1999,clasohm.ly:2006}.

All these suggest a different line to attack that may throw new light on matter of the surface potential and electrostatic boundary conditions. Indeed, if these effects exist, the ions of an electrolyte solution will concentrate in a very thin layer at the charged electrode by keeping their ability to diffuse. The need to invoke specific chemical adsorption is thereby removed, but we are unaware of previous work that has addressed the question of surface potential calculations in such a situation.
The only exception is probably a recent study~\cite{uematsu.y:2018}, where solutions of the Poisson-Boltzmann equation for $\Phi_s$ in terms of Jacobian elliptic functions  have been proposed. However, this has been done using the (Neumann) boundary condition of a fixed charge density $\sigma_0$ that is the case of insulators, and the closed-form analytic expressions have been derived only for highly dilute electrolyte solutions  in two opposite limits, of $\sigma \to \infty$ and of small $\Phi_0$.

Our purpose in this paper is to predict  $\Phi_s$ of conductors from known $\Phi_0$ by allowing for an explicit  Stern layer.
We develop an analytical theory based on the assumption that there is no specific adsorption and that the ions of the Stern layer are diffuse, but take into account that its permittivity is much smaller than the bulk one.  The model is simple enough to allow us to derive analytical approximations for the surface potential even when the ``wall'' potential is quite large. We show that on increasing $\Phi_0$ (from zero), the surface potential first grows linearly. On increasing the wall potential further, $\Phi_s$ augments nonlinearly and then saturates to a constant value. Such a scenario is universally valid, but the magnitude of $\Phi_s$ is also to a large extend controlled by the other parameters, such as permittivity contrast, which reduces $\Phi_s$, and  $\delta/\lambda$. When this ratio is rather large (in concentrated solutions), only a low $\Phi_s$ could be expected. However, when $\delta/\lambda$ is very small (highly dilute solutions), the surface potential can become quite large.
Our results indicate that $\Phi_s$ of conductors in model calculations with varying salt concentration  cannot be treated as fixed. However, in some conditions the conductor surface becomes of the fixed charge, which would normally be expected for insulators.

\section{Model}

We consider a planar wall of a fixed potential $\Phi_0$ located at $z=0$ and unbounded in the $x$ and $y$ directions. The wall is in contact with a reservoir of  1:1 salt solution at temperature $T$ and number density $n_{\infty}$, and treated as a perfectly polarizable electrode (\emph{i.e.}, no current crosses the conductor-electrolyte interface). The Debye screening length of a bulk solution, $\lambda=\left( 8\pi \ell _{B}n_{\infty}\right) ^{-1/2}$, is defined as usually with the Bjerrum
length, $\ell _{B}=\dfrac{e^{2}}{\varepsilon k_{B}T}$, where $e$ is the elementary positive charge, $k_{B}$ is the Boltzmann
constant, and $\varepsilon$ is the solvent permittivity.  By analyzing the experimental data it is more convenient to use the concentration $c_{\infty}[\rm{mol/l}]$, which is related to $n_{\infty} [\rm{m^{-3}}] $
as $n_{\infty} \simeq N_A \times 10^3 \times c_{\infty}$, where $N_A$ is Avogadro's number. The Bjerrum
length of water at $T \simeq 298$ K is equal to about $0.7$ nm leading to a
useful formula for 1:1 electrolyte
\begin{equation}\label{eq:DLength}
  \lambda [\rm{nm}] \simeq \frac{0.305 [\rm{nm}]}{\sqrt{c_{\infty}[\rm{mol/l}]} }
\end{equation}
Thus upon increasing $c_{\infty}$ from $10^{-6}$ to $10^{-1}$ mol/l the screening length is reduced from about 300 down to 1 nm.

\begin{figure}[t]
	\begin{center}
		\includegraphics[width=0.8\columnwidth]{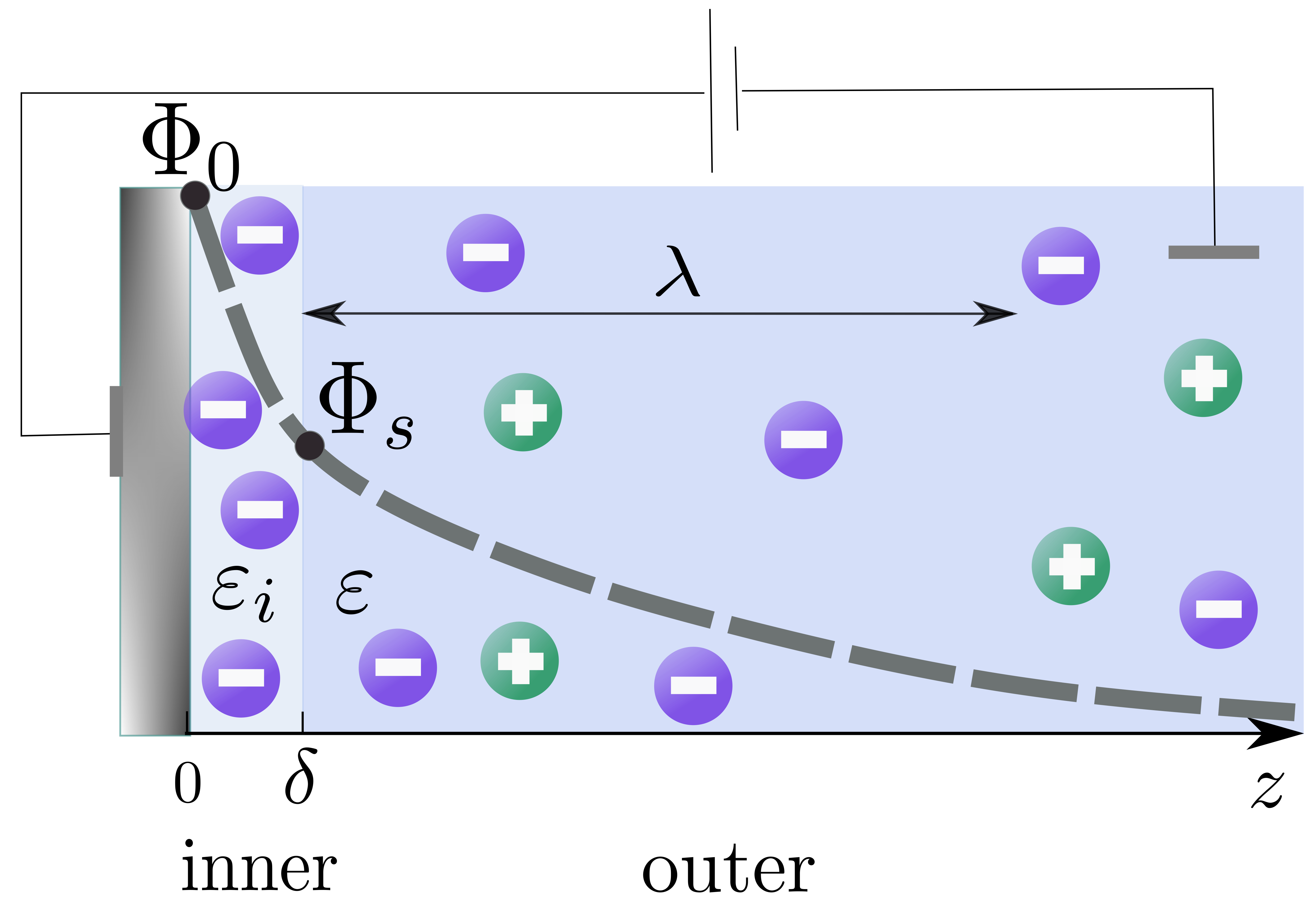}\\
	\end{center}
	\caption{Sketch of the perfectly polarisable electrode immersed in an electrolyte solution. An applied  potential $\Phi_0$ decays down to zero in the double electric layer. Its inner diffuse part is of thickness $\delta$ and permittivity $\varepsilon_i$. The  outer diffuse layer  extends up to distances of the order of the Debye length $\lambda$ from the inner layer. The surface potential $\Phi_s$, which is established self-consistently, is defined at their boundary.  }
	\label{fig:sketch}
\end{figure}

The system geometry is depicted on
Fig.~\ref{fig:sketch}. The electrolyte ions occupy the positive half
plane. Near a wall the double layer, which includes an adjacent to a body inner Stern layer and an outer electrostatic diffuse layer, is formed.
We assume that the inner layer of thickness $\delta$ has a reduced permittivity $\varepsilon_i$, and simultaneously exclude the traditional assumption that potential-determining ions are adsorbed and immobile, by allowing them to participate in the thermal motion. In other words, ions in both layers obey Boltzmann distribution, $c_{\pm }(z)=c_{\infty}\exp (\mp \phi_{i,o} (z))$, where $\phi_{i,o} (z)=e\Phi_{i,o}(z)/(k_{B}T)$ [recall that $k_{B}T/e \simeq 25$ mV] is the dimensionless electrostatic
potential,  the upper (lower) sign
corresponds to cations (anions), and index \emph{i} (\emph{o}) refers to an inner (outer) region.

Our aim is to calculate the surface potential $\phi_s$ that is established self-consistently at a distance $z=\delta$ from a wall. Clearly, there are still some uncertainties about what value to assign to $\delta$. Since inorganic ions have hydrodynamic diameters from 0.2 to 0.6  nm~\cite{kadhim.mj:2020} we might argue that in our model $\delta$  cannot be smaller. On the other hand, we identified the Stern layer as that of a reduced permittivity, and the best fit on experimental data  in pure water yielded $\delta \simeq 0.75 \pm 0.15$ nm~\cite{fumagalli.l:2018}. Taking into account these values, in our calculations below we will employ $\delta = 0.5$ nm, by giving only a few examples with smaller $\delta$. Note that exactly these values have been used before to interpret the measurements (conducted with mercury and other polarizable electrodes, \emph{i.e.} gold and carbon) of the capacity of the inner layer~\cite{conway.be:1999}.

The electrostatic potential satisfies the nonlinear Poisson-Boltzmann
equation
\begin{equation}\label{eq:PB2}
     \phi''_{i,o} = \lambda^{-2}_{i,o} \sinh \phi_{i,o}.
      \end{equation}
where $^{\prime}$ denotes $d/d z$. Recall that  the inner region is defined at $0\leq z \leq \delta$, an outer region at $z \geq \delta$, and $\lambda_o \equiv \lambda$. The inner Debye length $\lambda_i=\left( 8\pi \ell_i  n_{\infty}\right) ^{-1/2}$, where $\ell_{i}=\dfrac{e^2}{\varepsilon_i k_B T}$. In what follows  $\lambda^2 = \gamma \lambda_i^2$, where $\gamma = \varepsilon/\varepsilon_i$. Clearly, the inner Debye length is smaller than the outer one or equal to it. In principle, $\gamma$ can take any value from 1 to 80, but we focus here more on $\gamma \gg 1$. Note that the permittivity in the Stern layer has been earlier assumed, without adequate foundation, to have a value of 4~\cite{conway.be:1951}, and that consequent indirect experiments  suggested that $\varepsilon_i$ varies rather from 6 to 30~\cite{butt.hj:2003}, which corresponds to a reduction of  $\gamma$ from \emph{ca.} $13$ down to 3, but recent work reported $\varepsilon_i \simeq 2$ or $\gamma \simeq 40$~\cite{fumagalli.l:2018}.

The boundary condition at the conducting ``wall'' is that of a constant potential
\begin{equation}
\phi (0) = \phi_0.  \label{bc0}
\end{equation}%
We remark that $\phi_0$ can be quite large, so here we will use $\phi_0 \leq 40$ that corresponds to 1 V.

 The potential at the surface is continuous,%
\begin{equation*}
\phi _{s}=\phi_i (\delta)=\phi_o (\delta),
\end{equation*}%
but the gradient of potential changes across the surface as
\begin{equation}
\phi_i'(\delta)=\gamma\phi_o' (\delta).  \label{df}
\end{equation}

\section{General considerations and theory}
The outer solution of non-linear Eq.~\eqref{eq:PB2} is known~\cite{andelman.d:2006book}
 \begin{equation}\label{eq:PBSWexact}
\phi_o = 4 \artanh \left[e^{-(z-\delta)/\lambda}\tanh \left(\dfrac{\phi_s}{4}\right)\right],
\end{equation}
whence it follows that the outer derivative at the surface
 \begin{equation}\label{eq:der_delta}
\phi_o'(\delta) = -\frac{4}{\lambda} \dfrac{\tanh\dfrac{\phi_s}{4}}{1-\tanh^2\dfrac{\phi_s}{4}} = -\frac{2}{\lambda}\sinh\dfrac{\phi_s}{2}.
\end{equation}

Using \eqref{df} we then immediately obtain
 \begin{equation}\label{eq:difi}
\phi_i'(\delta) = -\frac{2 \gamma}{\lambda}\sinh\dfrac{\phi_s}{2}.
\end{equation}

We have to find a solution of non-linear equation \eqref{eq:PB2} in the inner region. Multiplying both parts by $\phi'_i$ we find
\begin{equation}
\frac{1}{2} \frac{d}{dz} \left(\dfrac{d\phi_i}{dz}\right)^{2} = \frac{\gamma}{\lambda^2} \sinh \phi_{i}\dfrac{d\phi_i}{dz}.
\end{equation}
This equation can be rewritten as
\begin{equation}
d \left(\dfrac{d\phi_i}{dz}\right)^{2} = \frac{2\gamma}{\lambda^2}  d \cosh \phi_{i}
\end{equation}

The first integration from $z$ to $\delta$ leads to
\begin{equation}\label{eq:difisquare}
\left(\phi_i'(z)\right)^{2} =\frac{2\gamma}{\lambda^2}  (\cosh \phi_{i} - \cosh \phi_{s})+\left(\phi_i'(\delta)\right)^{2}.
\end{equation}
It follows from \eqref{eq:difi} that
 \begin{equation}
(\phi_i'(\delta))^2 =  \frac{2 \gamma^2}{\lambda^2} (\cosh \phi_s - 1).
\end{equation}
Substituting this relation to \eqref{eq:difisquare}, we obtain
\begin{equation}\label{eq:finalDE}
\left(\dfrac{d\phi_i}{dz}\right)^{2} = \frac{2 \gamma^2}{ \lambda^{2}} \left[ \frac{\cosh \phi_{i} - \cosh \phi_{s}}{\gamma}  + \cosh \phi_s -1\right]
\end{equation}
or
\begin{equation}\label{eq:finalDE2}
\dfrac{d\phi_i}{dz} = \pm \frac{\sqrt{\gamma}}{ \lambda}\sqrt{2 \left[\cosh \phi_{i} + (\gamma - 1) \cosh \phi_s -\gamma\right]}
\end{equation}
The choice of sign in \eqref{eq:finalDE2} depends on whether $\phi_i$ is a monotonically increasing or decreasing function of $z$. Since we consider here positive $\phi_0$ the inner potential decays, the minus sign must be taken.

A further definite integration gives the equation
\begin{equation}\label{eq:final_int}
\int_{\phi_s}^{\phi_0}\dfrac{d\phi_i}{\sqrt{2\gamma \left[\cosh \phi_{i} + (\gamma - 1) \cosh \phi_s -\gamma\right]}} = \frac{ \delta}{ \lambda}
\end{equation}
that can be used to compute $\phi_0$ by setting a value of
surface potential, rather than the other way around. The right-hand side here is the ratio of the Stern layer thickness to the Debye length, both of which are given, so already at this stage it is clear that $\phi_s$ will always be expressed in terms
of $\delta/\lambda$. If we keep $\delta = 0.5$ nm fixed and vary $c_{\infty}$ in the range specified above, the value of  $\delta/\lambda$ is confined between ca. $2\times 10^{-3}$ to 0.5. For instance, if we set  $\lambda=30$ nm, then $\delta/\lambda \simeq 0.017$. If, say, $\gamma = 10$, then the potential
$\phi_s = 6$ is attained when $\phi_0 \simeq 11$. However, $\phi_s = 8$ cannot be attained since with this value of surface potential left-hand side is bounded by ca. 0.008, so that \eqref{eq:final_int} has no root. For completeness we mention that
\begin{equation}
\phi_i' (0) = -\frac{\sqrt{\gamma}}{ \lambda}\sqrt{2 \left[\cosh \phi_{0} + (\gamma - 1) \cosh \phi_s -\gamma\right]}
\end{equation}
and recall that $\phi_i' (\delta)$ is given by \eqref{eq:difi}.
For the above example of $\phi_s = 6$, we find $\phi_i' (0) \simeq -27$ nm$^{-1}$ and $\phi_i' (\delta) \simeq -7$ nm$^{-1}$. We see that $\phi_i' (0)$ exceeds $\phi_i' (\delta)$ which implies that the Stern layer is enriched by anions. Consequently, $\phi_i$ should be a concave function of $z$.

Note that if $\gamma = 1$ (no permittivity contrast), an exact analytical solution of \eqref{eq:final_int} can be found. Performing the integration we find that in this case the surface potential (formally defined at $z=\delta$) is given by
\begin{equation}\label{eq:gamma1}
\phi_s = 4 \artanh\left(\tanh \frac{\phi_0}{4} e^{-\delta/\lambda}\right),
\end{equation}
which turns to $\phi_s = 4 \artanh\left(e^{-\delta/\lambda}\right)$ at large $\phi_0$. For realistic values of $\delta/\lambda$ the latter would be anomalously large compared to experiments  and thus further supports the reduced permittivity concept.

\begin{figure}[h]
\centering
\includegraphics[width=0.7\columnwidth ]{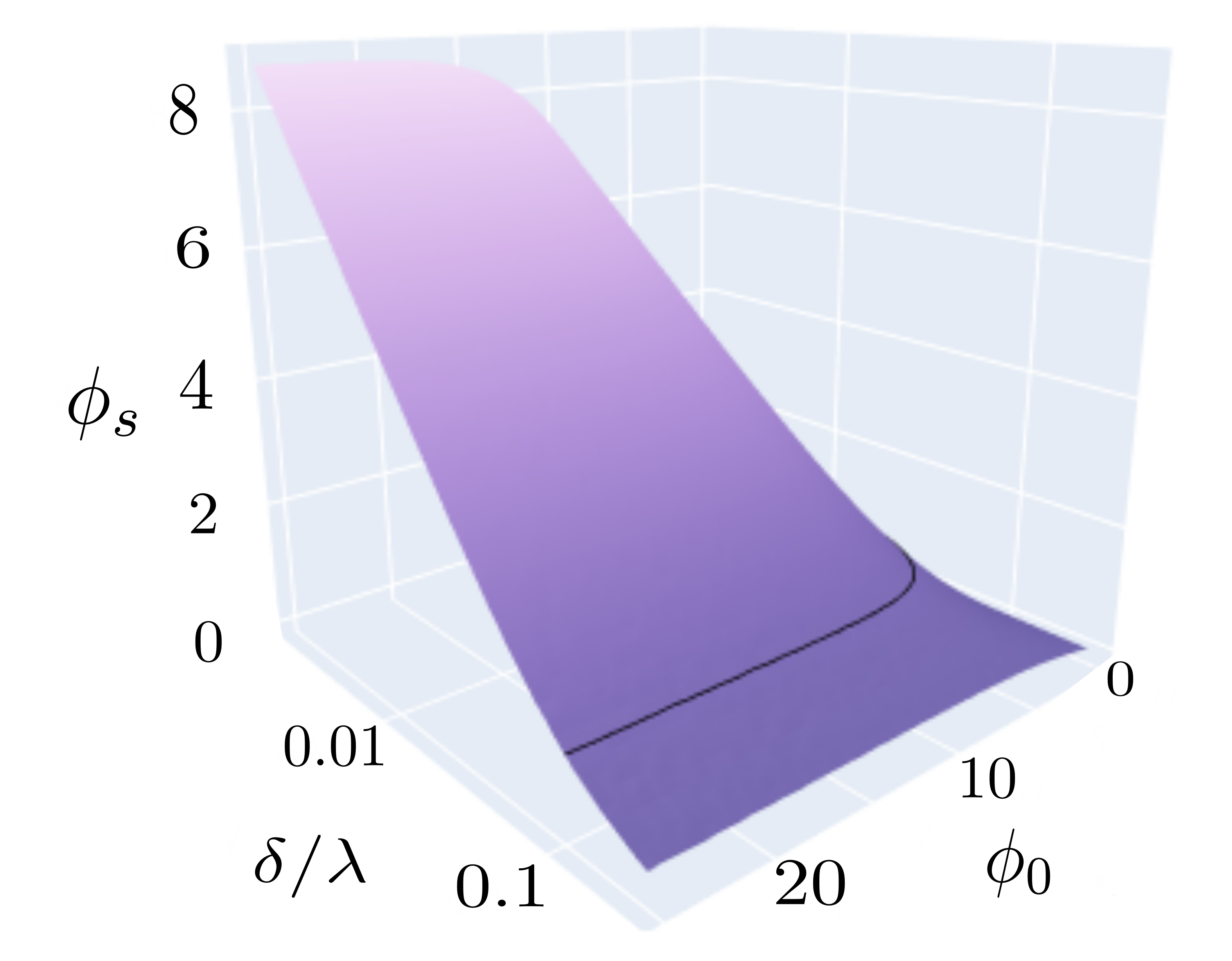}
\caption{Surface potential as a function of $\phi_0$ and $\delta/\lambda$ computed using $\gamma = 40$. A contour line connects the points, where $\phi_s = 1$.}
\label{fig:3d}
\end{figure}

Equation  \eqref{eq:final_int} is exact and applicable for any $\phi_i$, as well as being very well suited for numerical work.
In Fig.~\ref{fig:3d} we plot $\phi_s$ computed from Eq.~\eqref{eq:final_int} for $\gamma = 40$  as a function of two variables, $\phi_0$ and $\delta/\lambda$. An overall conclusion from this three dimensional plot is that $\phi_s$ augments on increasing $\phi_0$ when the latter is small enough, but for sufficiently large $\phi_0$ the surface potential attains its upper possible value and becomes insensitive to further increase in the intrinsic potential. It can also be seen that $\phi_s$
is extremely sensitive  to $\delta/\lambda$. If we keep $\phi_0$ and $\delta$ fixed, then on increasing $\lambda$ (decreasing the bulk concentration of salt), the magnitude of $\phi_s$ augments. The region of  $\phi_s \leq 1$  exists for $\delta/\lambda$ of the order of $10^{-1}$, even if $\phi_0$ is high. The high surface potential can only be  attained in more dilute solutions, where $\delta/\lambda = O(10^{-2})$ and smaller.

It is of considerable  interest of yielding useful (approximate) analytical results  describing the general picture or its particular regions shown in Fig.~\ref{fig:3d}, but
an analytic solution of \eqref{eq:final_int} is straightforward only when $\gamma = 1$ (see Eq.~\eqref{eq:gamma1}).
The case of $\gamma\neq 1$ we are interested in is challenging since the solution of  \eqref{eq:final_int} could be expressed only in terms of elliptic functions. This, in turn, would require the subsequent numerical solution of complex transcendental equations. However, in the limits of small and large surface potentials approximate analytical results can be obtained. Below we consider these two distinct limits.

\section{Surface potential}
\subsection{Small surface potentials}

We focus first to the case of a small surface potential, $\phi_s \leq 1$. Such a situation is expected in the very large range of $\delta/\lambda$ and $\phi_0$ as discussed above (and well seen in Fig.~\ref{fig:3d}).
Expanding \eqref{eq:PBSWexact} in the Taylor series about $\phi_s = 0$ leads to a familiar equation for an outer region
\begin{equation}
  \phi_o \simeq \phi_s e^{-(z-\delta)/\lambda}.
\end{equation}

 If we do not make any assumptions about $\phi_0$, in the inner region Eq.~\eqref{eq:final_int} can be expanded similarly to give  the following approximation:
\begin{equation}\label{eq:finalDElinear_new}
\int_{\phi_s}^{\phi_0} \dfrac{d\phi_i}{\sqrt{2 (\cosh \phi_{i}-1)  +  \phi_s^2 (\gamma - 1)}} \simeq \frac{\sqrt{ \gamma} \delta}{ \lambda}
\end{equation}
or
\begin{equation}\label{eq:finalDElinear_new3}
-i \mathcal{F} \left(\frac{2 i \sinh \left(\dfrac{\phi_{i}}{2}\right)}{\phi_s \sqrt{\gamma - 1}} , \dfrac{\phi_s \sqrt{\gamma - 1}}{2}\right)\bigg|^{\phi_0}_{\phi_s}\simeq \frac{\sqrt{ \gamma} \delta}{ \lambda},
\end{equation}
where $i^2 = -1$ and $\mathcal{F}=\mathrm{sn}^{-1}$ is incomplete elliptic integral of the first kind.

\begin{figure}[h]
\centering
\includegraphics[width=0.7\columnwidth]{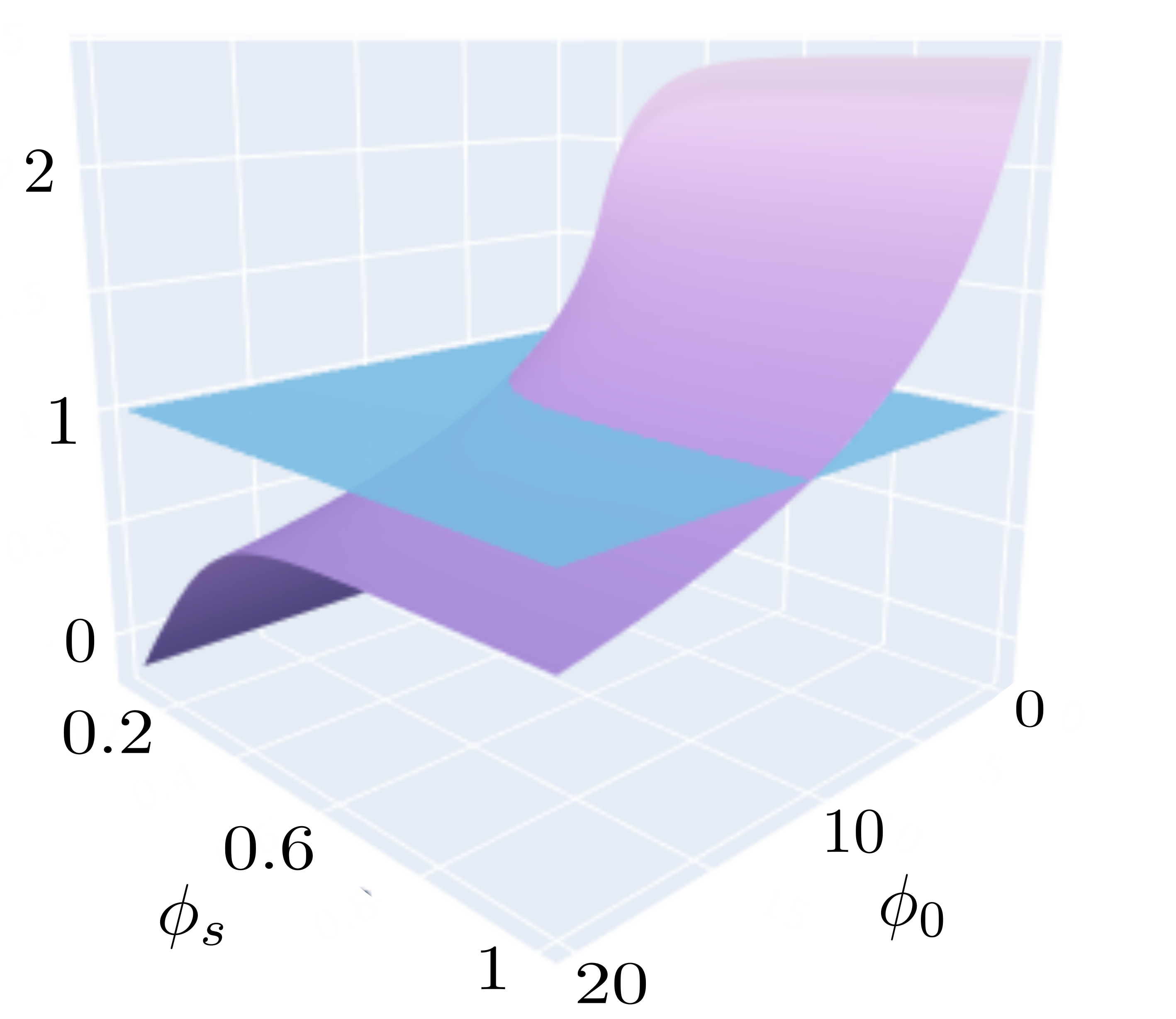}\\
\caption{Definite integral (l.h.s. in Eq.~\eqref{eq:finalDElinear_new3}) computed using $\gamma=40$ and
plotted as a function of $\phi_s$ and $\phi_0$ along with a plane that corresponds to r.h.s.=1}
\label{fig:test}
\end{figure}

This case is illustrated in Fig.~\ref{fig:test}, where we plot left-hand side of Eq.~\eqref{eq:finalDElinear_new3} as a function of $\phi_0$ and (small) $\phi_s$ using $\gamma=40$. Also included is its right-hand side (a plane) taken as equal to unity for this specimen example. With a given value of $\gamma$ this plane corresponds to $\delta/\lambda = 1/\sqrt{40} \simeq 0.16$. If $\delta = 0.5$ nm, this value would be attained with $\lambda \simeq 3$ nm, i.e. when $c_{\infty} \simeq 10^{-2}$ mol/l. A cross-over of the two surfaces in Fig.~\ref{fig:test} corresponds to the $\phi_s$ \emph{vs.} $\phi_0$ curve at this salt concentration.

\begin{figure}[h]
\centering
\includegraphics[width=0.99\columnwidth ]{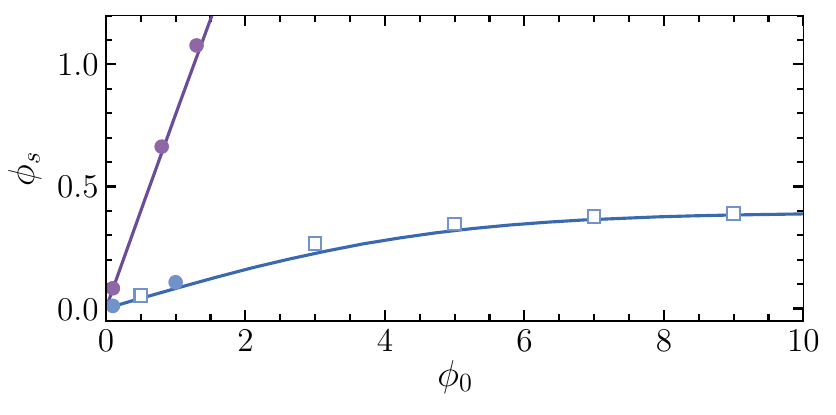}
\caption{Surface potential $\phi_{s}$ \emph{vs.} $\phi_0$ computed using $\delta = 0.5$ nm and $\gamma = 40$ at $c_{\infty} = 10^{-5}$ mol/l (upper solid line) and $10^{-2}$ mol/l (lower solid curve). Filled circles correspond to Eq.~\eqref{eq:phis_small_general}. Open squares are calculations from Eq.~\eqref{eq:small_pot}.}
\label{fig:small_pot1}
\end{figure}

Figure~\ref{fig:small_pot1} displays such a curve computed with the same parameters as in Fig.~\ref{fig:test}. Also included is the low surface potential region of the  $\phi_s$ \emph{vs.} $\phi_0$ curve computed using the same $\delta$ and $\gamma$, but for  $c_{\infty} = 10^{-5}$ mol/l (or $\delta/\lambda \simeq 5 \times 10^{-3}$).  It can be seen that for small $\phi_0$ the surface potential responds (augments) linearly and that the slope of this line is much larger at lower salt concentration. As a result, the uppermost part of the curve computed for a dilute solution escapes from this plot. However, calculations using $c_{\infty} = 10^{-2}$ mol/l yield  small $\phi_s$ for any intrinsic potentials, and we see that $\phi_s$ tends to some upper value and becomes independent on $\phi_0$ when the latter is sufficiently large.

The physical interpreting of these results requires the general expansion of the integrand in \eqref{eq:finalDElinear_new} that would be valid in the whole interval. This  is a challenging if not impossible task. Thus below we consider two special cases only, namely, of low and high $\phi_0$.

\subsubsection{Small intrinsic potential }

In the limit of small ``wall'' potentials, $\phi_0 \leq 1$, the expression for $\phi_{i}$ can always be found by solving the linearised Poisson-Boltzmann equation, but here we derive it by using the above general results.
In the inner region Eq.~\eqref{eq:finalDElinear_new} can be further expanded about $\phi_i$, and we obtain
\begin{equation}\label{eq:finalDElinear}
\dfrac{-d\phi_i}{\sqrt{\phi_{i}^2  +  \phi_s^2 (\gamma - 1)}} \simeq \frac{\sqrt{ \gamma}}{ \lambda} dz
\end{equation}
Integrating l.h.s. from $\phi_0$ to $\phi_s$ and r.h.s. from 0 to $\delta$ one can derive
\begin{equation}\label{eq:smfi}
\ln \left( \frac{\phi _{0}+\sqrt{\phi _{0}^{2}+(\gamma -1)\phi
_{s}^{2}}}{\left( 1+\sqrt{\gamma }\right) \phi _{s}}\right) \simeq \frac{\sqrt{ \gamma}\delta}{ \lambda},
\end{equation}%
which yields

\begin{equation}
\phi_{s} \simeq \dfrac{2 \phi_0 e^{\frac{\sqrt{ \gamma}}{ \lambda} \delta}(1+\sqrt{\gamma})}{(1+\sqrt{\gamma})^2 e^{\frac{2\sqrt{ \gamma}}{ \lambda} \delta}+ 1 - \gamma}.
\end{equation}
The last equation may be reexpressed as
\begin{equation}\label{eq:phis_small_general}
   \phi_s \simeq \dfrac{\phi_0}{\cosh \left(\frac{\sqrt{ \gamma}}{ \lambda} \delta \right) + \sqrt{ \gamma} \sinh \left(\frac{\sqrt{ \gamma}}{ \lambda} \delta \right)}.
  \end{equation}
Thus in this limit $\phi_s$ increases linearly with $\phi_0$. The slope of this line is defined by $\delta/\lambda$ and also depends on $\sqrt{\gamma}$. The results of calculations from \eqref{eq:phis_small_general} are included in Fig.~\ref{fig:small_pot1} and it is seen that they nearly coincide with the numerical data.

We remark and stress that although we applied the constant potential condition at the ``wall'', the value of $\phi_s$ depends on $\lambda$ (salt concentration), which implies that in the general case this classical  boundary condition cannot be imposed at the surface of conductors immersed in electrolyte solutions.
The only exception is a situation of
highly dilute solutions (of a very large $\lambda$), where the argument $\frac{\sqrt{\gamma }}{\lambda }\delta $ becomes small and \eqref{eq:phis_small_general} can be reduced to
\begin{equation}\label{eq:uematsu}
   \phi_s \simeq \dfrac{\phi_0}{1 + \dfrac{\gamma \delta}{ \lambda} } \simeq \phi_0 \left(1- \frac{\gamma \delta}{ \lambda}  \right),
  \end{equation}
which implies that the drop of an inner potential is negligibly small. Later we shall see that this is the only possible situation where $\phi_s \simeq \phi_0$ and, hence, the fixed surface potential condition is justified.
Estimating the order of magnitudes one can argue the first equation in \eqref{eq:uematsu} is applicable when $\delta/\lambda \leq 0.1$, while the second relation, which is identical to that derived by \citet{uematsu.y:2018}, is justified provided $\delta/\lambda = O(10^{-3})$. Generally, in the linear mode $\phi_s$  obeys Eq.~\eqref{eq:phis_small_general} that is valid at any $\delta/\lambda$. Note, however, that if $\gamma$ is large, then at sufficiently large $\delta/\lambda$ the surface potential becomes quite small. Indeed, when $\frac{\sqrt{ \gamma} \delta}{ \lambda} \geq 1$, Eq.~\eqref{eq:phis_small_general} can be approximated by
\begin{equation}\label{eq:phis_small_general2}
   \phi_s \simeq \dfrac{2 \phi_0 e^{-\sqrt{ \gamma} \delta/\lambda}}{1 + \sqrt{ \gamma}}.
  \end{equation}

\subsubsection{Large intrinsic potential }

When $\phi_0$ is finite the expansion \eqref{eq:finalDElinear} is valid only near the surface, where $\phi_i = O(\phi_s)$, while
for $\phi _{i}\gg \phi_s$ the integrand in \eqref{eq:finalDElinear_new} can be approximated by $1/\sqrt{2(\cosh \phi _{i}-1)}$.
Dividing up the range of the integral in Eq.~\eqref{eq:finalDElinear_new} into appropriate parts
\begin{equation}
 \int_{\phi _{s}}^{\phi^{\star}}\dfrac{d \phi_i}{\sqrt{\phi _{i}^{2}+(\gamma -1)\phi
_{s}^{2}}} + \int_{\phi^{\star}}^{\phi _{0}}\dfrac{d\phi _{i}}{2\sinh\left(\dfrac{\phi _{i}}{2}\right)} \simeq \frac{\sqrt{ \gamma}\delta}{ \lambda},
\end{equation}
and performing the integration we obtain
\begin{equation}
 \ln \left( \frac{\phi^{\star}+\sqrt{\phi^\star{^2}+(\gamma -1)\phi_{s}^2}}{\left( 1+\sqrt{\gamma }\right) \phi_{s}} \times \frac{4 \tanh \frac{\phi_{0}}{4}}{\phi^{\star}}\right)\simeq \frac{\sqrt{ \gamma}\delta}{ \lambda},
\end{equation}
which can be simplified to
\begin{equation}
  \ln \left( \frac{8}{\left( 1+\sqrt{\gamma }\right) \phi _{s}}\tanh
\dfrac{\phi _{0}}{4}\right)\simeq \frac{\sqrt{ \gamma}\delta}{ \lambda}
\end{equation}
provided $\phi^{\star}$ is large compared to $\phi_s \sqrt{\gamma - 1}$.

Thus we might argue that to leading order the surface potential can be approximated by
\begin{equation}\label{eq:small_pot}
\phi _{s}\simeq\frac{8}{1+\sqrt{\gamma }}\tanh \dfrac{\phi _{0}}{4}e^{-\frac{\sqrt{%
\gamma }}{\lambda }\delta }
\end{equation}
This equation is applicable only if $\frac{\sqrt{\gamma }}{\lambda }\delta $ is large enough to provide $\phi_s \leq 1$, but one can speculate that it could be valid for any $\phi_0$. Indeed, at small intrinsic potential this equation reduces to \eqref{eq:phis_small_general2}.
If $\phi_0 \geq 4$, Eq.~\eqref{eq:small_pot} yields the upper bound of $\phi_s$ that can be attained
\begin{equation}\label{eq:lowphisat}
\phi _{s}\simeq \frac{8}{1+\sqrt{\gamma }}e^{-\frac{\sqrt{%
\gamma }}{\lambda }\delta }
\end{equation}
This value corresponds to the surface potential at the plateau region.

The calculations from Eq.~\eqref{eq:small_pot} are included in Fig.~\ref{fig:small_pot1}. It can be seen that the fit of the lower $\phi_s$-curve that corresponds to $c_{\infty} = 10^{-2}$ mol/l (and, consequently, $\frac{\sqrt{\gamma }}{\lambda }\delta  \simeq 1.1$) is indeed very good for all $\phi_0$. It is important to recall that Eq.~\eqref{eq:small_pot} is inapplicable for an upper curve in Fig.~\ref{fig:small_pot1}, which also contains a plateau branch, but at large $\phi_s$. Below we shall see that the when the surface potential is large, the approximations of general Eq.~\eqref{eq:final_int} become very different.

\subsection{Large surface potentials}

\begin{figure}[h]
\centering
\includegraphics[width=0.99\columnwidth ]{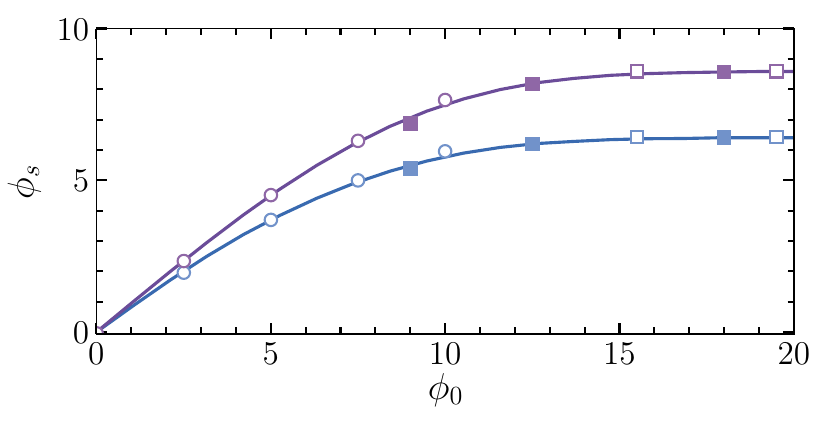}
\caption{Surface potential $\phi_{s}$ \emph{vs.} $\phi_0$ computed using $\gamma = 10$ (top), $\gamma = 40$ (bottom) with fixed $\delta = 0.5$ nm,  $c_{\infty} = 10^{-5}$ mol/l. Circles are calculations from Eq.~\eqref{eq:lambert}. Filled and open squares correspond to Eqs.~\eqref{eq:large_pot} and \eqref{eq:plateau}}
\label{fig:large_pot}
\end{figure}

If we increase the intrinsic potential of conductors in very dilute solutions (here $c_{\infty} = 10^{-5}$ mol/l), we move to the situation described by Fig.~\ref{fig:large_pot}. The calculations are made using $\delta = 0.5$ nm, and two values of $\gamma$. We see that on increasing $\gamma$ the magnitude of $\phi_s$ decreases. For both $\gamma$, the surface potential first shows an increase with $\phi_0$ and then saturates. Thus, qualitatively the picture is the same as we observed for $c_{\infty} = 10^{-2}$ mol/l (see Fig.~\ref{fig:small_pot1}), but the values of $\phi_s$ are now large, which requires different analysis.

Provided $\phi_s \ge 3$, one can approximate $\cosh \phi_s - 1 \simeq \cosh \phi_s $. Taking into account that $2\cosh \phi_{i,s} \simeq e^{\phi_{i,s}}$, we might argue that the sensible approximation of \eqref{eq:finalDE2} should be
\begin{equation}\label{eq:deriv}
\dfrac{d\phi_i}{dz} \simeq - \frac{\sqrt{\gamma}}{ \lambda} \sqrt{e^{\phi_{i}}+ (\gamma-1) e^{\phi_{s}}}
\end{equation}

Equation \eqref{eq:deriv} can be rewritten as
\begin{equation}
\dfrac{-d\phi_i}{\sqrt{e^{\phi_{i}}+\mathcal{S}}}  \simeq  \frac{\sqrt{\gamma}}{ \lambda} dz,
\end{equation}
where we introduce the function $\mathcal{S} =  e^{\phi_{s}}(\gamma - 1)$.
Integrating from 0 to $\delta$ one can derive the transcendental relation
\begin{equation}\label{eq:result2}
2 \arcoth \sqrt{1+\dfrac{e^{\phi_{s}}}{\mathcal{S} }}  - 2 \arcoth \sqrt{1+\dfrac{e^{\phi_{0}}}{\mathcal{S} }} =  \frac{\delta\sqrt{\gamma \mathcal{S}}}{\lambda},
\end{equation}
which allows one to determine $\mathcal{S}$ and $\phi_s$ at a given $\phi_0$, but generally only numerically.

In the case of $\gamma \gg 1$, the value of $\mathcal{S}$ becomes very large and $e^{\phi_{s}}/S = 1/(\gamma -1) \ll 1$. One can then argue that the first term in \eqref{eq:result2} can be safely approximated as
\begin{equation}
 2 \arcoth \sqrt{1+\dfrac{e^{\phi_{s}}}{\mathcal{S} }}  \simeq \ln \dfrac{4 \mathcal{S}}{e^{\phi_s}} \simeq \ln [4 (\gamma - 1)]
\end{equation}
Therefore, Eq.~\eqref{eq:result2} can be reduced to
\begin{equation}\label{eq:LPgen}
\ln \dfrac{4 \mathcal{S}}{e^{\phi_s}} - \frac{\delta\sqrt{\gamma \phi_0}}{\lambda}\sqrt{\dfrac{\mathcal{S} }{e^{\phi_{0}}}}
  = 2 \arcoth \sqrt{1+\dfrac{e^{\phi_{0}}}{\mathcal{S} }}.
\end{equation}

As specified above, $e^{\phi_s}/\mathcal{S}$ is always small for $\gamma \gg 1$, but $e^{\phi_0}/\mathcal{S}$ can take any value. Below we consider the limits of small and large $e^{\phi_0}/\mathcal{S}$ separately.

\subsubsection{Small $e^{\phi_0}/\mathcal{S}$}

If $e^{\phi_0}/\mathcal{S} \ll 1$, Eq.~\eqref{eq:result2} can be expanded in Taylor series, and to leading order
\begin{equation}
\ln \dfrac{4 \mathcal{S}}{e^{\phi_s}} - \ln \dfrac{4 \mathcal{S}}{e^{\phi_0}} \simeq \frac{\delta\sqrt{\gamma \mathcal{S}}}{\lambda}
\end{equation}
Whence
\begin{equation}
\ln \dfrac{e^{\phi_0}}{e^{\phi_s}} \simeq \frac{\delta\sqrt{\gamma \mathcal{S}}}{\lambda}
\end{equation}
and
\begin{equation}\label{eq:phis_phi0}
\phi_{s} \simeq  \phi_{0} - \dfrac{\delta \sqrt{ \gamma (\gamma - 1)}}{\lambda} \sqrt{e^{\phi_{s}}}.
\end{equation}
The surface potential can be obtained by solving transcendental equation \eqref{eq:phis_phi0}, which gives
\begin{equation}\label{eq:lambert}
\phi_{s}\simeq \phi_{0} - 2 \mathcal{W} \left( \dfrac{ \delta }{2 \lambda} \sqrt{\gamma (\gamma - 1) e^{\phi_0}} \right).
\end{equation}
Here $\mathcal{W}$ is the Lambert W function (also called the omega function). The calculations from Eq.~\eqref{eq:lambert} are included in Fig.~\ref{fig:large_pot}, and we conclude that the fits are quite good.

Summarising, in this (intermediate) mode the surface potential shows a nonlinear growth with $\phi_0$. Another important conclusion is that  $\phi_s$ is again appears to be salt-dependent like it was in the linear mode.

\subsubsection{Large $e^{\phi_0}/\mathcal{S}$}

If $e^{\phi_0}/\mathcal{S} \gg 1,$ which is  the case of very high (applied) ``wall'' potentials, Eq.~\eqref{eq:LPgen} reduces to
\begin{equation}
\ln \dfrac{4 \mathcal{S}}{e^{\phi_s}}
  \simeq \sqrt{\dfrac{4\mathcal{S} }{e^{\phi_{0}}}} \left( 1 + \frac{\delta\sqrt{\gamma \phi_0}}{2\lambda}\right),
\end{equation}
which leads to
\begin{equation}\label{eq:large_pot}
 \phi_s \simeq \ln \left(  \dfrac{\ln^2 [4(\gamma-1)]}{\left( \dfrac{\delta\sqrt{\gamma }}{2\lambda}+\sqrt{e^{-\phi_{0}}}\right)^2 4(\gamma-1)}\right)
\end{equation}

Thus, on increasing $\phi_0$ the surface potential augments weakly logarithmically and then saturates to
\begin{equation}\label{eq:plateau}
\phi_{s} \simeq \ln \left( \dfrac{\lambda^2 \ln^2 [4(\gamma - 1)] }{\delta^2 \gamma (\gamma - 1)} \right).
\end{equation}

Calculations from Eqs.~\eqref{eq:large_pot} and \eqref{eq:plateau} are also included in Fig.~\ref{fig:large_pot} and fit very well the appropriate regions of the $\phi_s$ \emph{vs.} $\phi_0$ curves. Clearly, Eq.~\eqref{eq:plateau} describes an upper bound that constraints the attainable $\phi_s$. There has been a report on such a phenomenon by \citet{uematsu.y:2018} and it has been interpreted in this fashion. Their asymptotic equation derived in the limit of high surface charge and low salinity is constructed similarly to \eqref{eq:plateau}, but includes $4 \arcosh^2\sqrt{\gamma}$ instead of $\ln^2 [4(\gamma - 1)]$. If, however, we compare these two functions numerically it is evident that when $\gamma \gg 1$, they are very close.

\begin{figure}[h]
\centering
\includegraphics[width=0.99\columnwidth ]{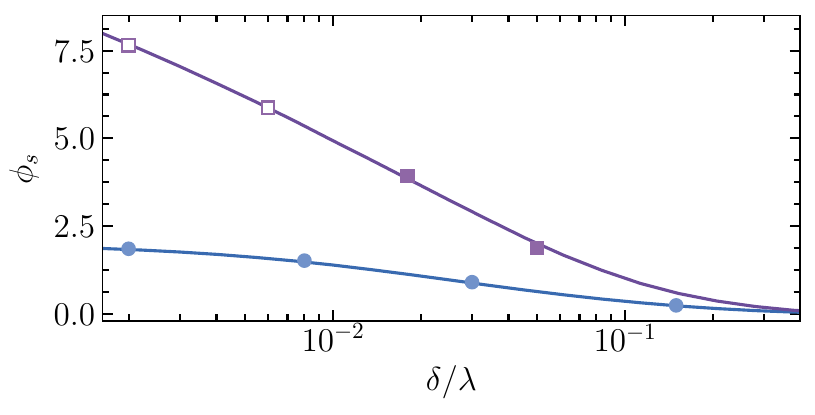}
\caption{Surface potential $\phi_{s}$ \emph{vs.} $\delta/\lambda$ computed using $\gamma = 40$ for $\phi_0 = 12$ (upper curve) and $\phi_0 = 2$ (lower curve). Circles correspond to Eqs.~\eqref{eq:phis_small_general}. Filled and open squares are calculations from Eqs.~\eqref{eq:large_pot} and \eqref{eq:plateau}}
\label{fig:delta}
\end{figure}

Equation~\eqref{eq:plateau} provides theoretical guidance for system optimisation since it can be tuned by adjusting salt concentration.  Say, if we set $\gamma=10$ and $\delta = 0.5$ nm, the decrease of $\lambda$ from 300 nm down to 10 nm would reduce attainable $\phi_s$ from 10.8 down to 4.
However, not all $\delta/\lambda$ are allowed. Since we assumed $\phi_{s} \geq 3$,
\begin{equation}\label{eq:limg}
\dfrac{\delta}{\lambda} \leq \dfrac{ \ln [4(\gamma - 1)] }{\sqrt{e^{3} \gamma (\gamma - 1)}}
\end{equation}
is required for \eqref{eq:plateau} to be applicable. For instance, at $\gamma = 10$ it is necessary to provide $\delta/\lambda \leq 0.084$. If $\delta=0.5$ nm, this would give minimal $\lambda \simeq 6$ nm, which corresponds to $c_{\infty}\leq 3\times 10^{-3}$ mol/l. To examine the significance of salt concentration more closely, in Fig.~\ref{fig:delta} we plot $\phi_s$ computed for $\phi_0 = 12$ and $\gamma = 40$ as a function of $\delta/\lambda \propto \sqrt{c_{\infty}}$.  It can be seen that at lower concentrations $\phi_s$ is large and reduces on increasing $\delta/\lambda$ being perfectly described first by Eq.~\eqref{eq:plateau} and then by \eqref{eq:large_pot}. We remark that at low concentrations the variation of $\phi_s$  appears as linear in this lin-log plot.
Also included in Fig.~\ref{fig:delta} is the $\phi_s$-curve calculated using $\phi_0 = 2$, which is intended to illustrate that in this case $\phi_s$ is much smaller and that the approximation~\eqref{eq:phis_small_general} has validity well
beyond the range of the original assumptions.

\begin{figure}[h]
\centering
\includegraphics[width=0.99\columnwidth ]{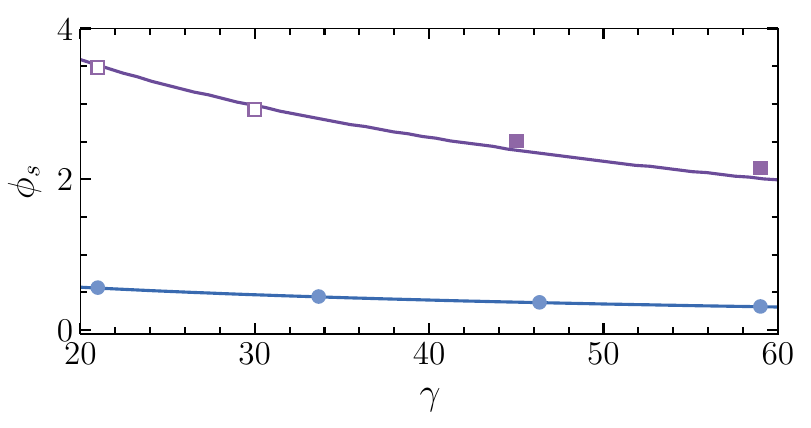}
\caption{Surface potential $\phi_{s}$ \emph{vs.} $\gamma$, computed using $\phi_0 = 12$ (top), $\phi_0 = 1$ (bottom) with fixed $\delta = 0.5$ nm,  $c_{\infty} = 5 \times 10^{-4}$ mol/l. Circles are obtained from Eq.~\eqref{eq:phis_small_general}. Filled and open squares are calculations from Eqs.~\eqref{eq:large_pot} and \eqref{eq:plateau}.}
\label{fig:gamma}
\end{figure}

Our previous discussion already showed that  the permittivity contrast $\gamma$ reduces the surface potential. This trend becomes clear in Fig.~\ref{fig:gamma}, where we keep $c_{\infty} = 5 \times 10^{-4}$ mol/l fixed and vary $\gamma$ from 20 to 60. It can be seen that $\phi_s$ decreases monotonically with $\gamma$. For reasons explained above it is appropriate to use different equations to fit left and right-side branches of the high $\phi_0$ curve. Say, with $\delta = 0.5$ nm and given salt concentration one can estimate that $\delta/\lambda \simeq 0.04$. From \eqref{eq:limg} we then conclude that Eq.~\eqref{eq:plateau} applies only if $\gamma \leq 30$. For larger $\gamma$ the surface potential has not reached a saturation yet, so that to provide a better fit to numerical data we use \eqref{eq:large_pot}.

\section{Toward tuning surface charge}

The analysis and examples described so far have been focussed on calculation of $\phi_s$.
We turn now to the surface charge density $\sigma$, which is related to $\phi_s$ as
\begin{equation}\label{eq:BC_charge}
\phi_o'(\delta) = - \dfrac{2}{\ell _{GC}},
\end{equation}
where we introduced the Gouy-Chapman length~\cite{andelman.d:2006book}
\begin{equation}\label{eq:LGC}
  \ell _{GC}=\dfrac{e}{2\pi \sigma \ell _{B}}
\end{equation}
that scales as the inverse of the surface charge. It follows then from Eq.~\eqref{eq:der_delta} that the relation between $\phi_s$ and $\ell _{GC}$ is given by
\begin{equation}\label{eq:Graham_new}
 \phi_s = 2 \arsinh \left(\dfrac{\lambda}{\ell _{GC}} \right),
\end{equation}
which is identical to the Grahame equation~\cite{andelman.d:2006book,israelachvili.jn:2011,vinogradova.oi:2024}. The ratio $\lambda/\ell _{GC}$ is proportional to the surface charge density and may be considered as an effective (dimensionless) surface charge.

\begin{figure}[h]
	\centering
	\includegraphics[width=0.99\columnwidth ]{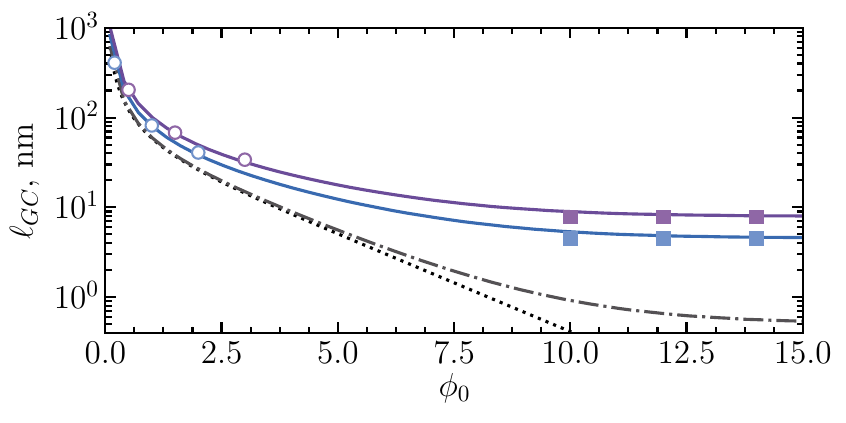}
	\caption{The Gouy-Chapman length $\ell_{GC}$ as a function of intrinsic potential $\phi_{0}$. Solid curves from top to bottom are computed for $\gamma = 40$ and $20$ using $\delta = 0.5$ nm and $c_{\infty} = 10^{-4}$~mol/l. Open circles and filled squares show predictions of Eqs.~\eqref{eq:small_phi0} and \eqref{eq:CC}.
 Dash-dotted curve indicates the case when $\gamma=1$ and obtained by substituting Eq.~\eqref{eq:gamma1} into Eq.~\eqref{eq:Graham_new}.  Dotted curve corresponds to Eq.~\eqref{eq:Graham_new} using $\phi_s = \phi_{0}$. }
	\label{fig:contact_th}
\end{figure}

The Gouy-Chapman lengths calculated from Eq.~\eqref{eq:Graham_new} and  plotted as a function of $\phi_0$ are given in Fig.~\ref{fig:contact_th}. The calculations
are made for a rather dilute solution of  $c_{\infty} = 10^{-4}$~mol/l ($\lambda \simeq 30$ nm) using $\delta = 0.5$ nm and three values of $\gamma$, namely, 40, 20, and 1. For $\gamma = 1$ the surface potential in Eq.~\eqref{eq:Graham_new} has been obtained from \eqref{eq:gamma1}, but for other values of $\gamma$ computed as in the previous section. The results presented in Fig.~\ref{fig:contact_th} show a decay of $\ell_{GC}$ on increasing $\phi_0$ from very high to much smaller and constant values at $\phi_0 \gg 1$, namely $\ell_{GC}\simeq 7.8, \, 4.5,$ and 0.5 nm. Thus on reducing $\gamma$ from 40 down to 1 we augment $\sigma$ from ca. 4.5 to 67 mC/m$^2$. The latter case corresponds to one unit charge per $ 2.4 \, \mathrm{nm^2}$, which appears too high, and thus further supports the concept of reduced permittivity. Note that if we would define $\phi_s$ at the ``wall'', \emph{i.e.} consider $\phi_s = \phi_ 0$, Eq.~\eqref{eq:Graham_new} would predict no saturation effect for $\ell_{GC}$, which implies that $\sigma$ always  grows with $\phi_0$. Such a scenario does not appear to be realistic due to a finite size of real ions.

The nature of curves in Fig.~\ref{fig:contact_th} obtained for large $\gamma$  is apparent. When $\lambda/\ell _{GC}$ is small, Eq.~\eqref{eq:Graham_new} can be simplified to
\begin{equation}\label{eq:Grahame_smallphi}
\phi_s   \simeq \dfrac{2 \lambda}{\ell _{GC}}.
\end{equation}
This implies that the surface potential is low, and for a branch of small $\phi_0$ is described by Eq.~\eqref{eq:phis_small_general}. Then
substituting $\phi_s$ given by Eq.~\eqref{eq:phis_small_general} into \eqref{eq:Grahame_smallphi} it is straightforward to derive
\begin{equation}\label{eq:small_phi0}
\ell _{GC} \simeq \dfrac{2 \lambda \left( \cosh \left(\frac{\sqrt{ \gamma}}{ \lambda} \delta \right) + \sqrt{ \gamma} \sinh \left(\frac{\sqrt{ \gamma}}{ \lambda} \delta \right)\right)}{\phi_0}
\end{equation}
In highly dilute solutions to leading order $\ell _{GC} \propto \lambda$, inversely proportional to (small) $\phi_0$ and does not depend on $\gamma$. Thus it can be very large and turns to $\infty$ when $\phi_0 \to 0$. A calculation of the Gouy-Chapman length from \eqref{eq:small_phi0} is included to Fig.~\ref{fig:contact_th} to illustrate that the fit is excellent and we conclude that the expression \eqref{eq:small_phi0} is applicable even beyond the range of initial assumptions (here - at finite intrinsic potentials of a conductor, up to $\phi_0\simeq 3$).

For large $\lambda/\ell _{GC}$ the Grahame equation \eqref{eq:Graham_new} can be approximated as
\begin{equation}
  \phi_s  \simeq 2 \ln \left(\dfrac{2\lambda}{\ell _{GC}} \right).
\end{equation}
For the plateau branch of Fig.~\ref{fig:contact_th} we can estimate $\phi_s \simeq 4$ for $\gamma=40$ and $\phi_s \simeq 5$ when $\gamma=20$. Both  values are large, and it is easy to check that $e^{\phi_0}/\mathcal{S}$ is large too. Thus, $\phi_s$ obeys \eqref{eq:plateau}, which immediately yields
\begin{equation}\label{eq:CC}
  \ell _{GC} \simeq \dfrac{2 \delta \gamma}{\ln [4 (\gamma - 1)]}\sqrt{\dfrac{\gamma - 1}{ \gamma }}.
\end{equation}
This equation describes the Gouy-Chapman length at saturation, which can be seen as a lower bound on $\ell _{GC}$ at given conditions. We see that $\ell _{GC}$ does neither depend on $\lambda$ nor $\phi_0$, but depends on $\delta$ and $\gamma$. Since $ \ell _{GC} \propto \delta$, it should be rather small. The calculations from Eq.~\eqref{eq:CC} fit numerical data in Fig.~\ref{fig:contact_th} very well.

We stress that Eq.~\eqref{eq:plateau}, and consequently \eqref{eq:CC} too, is justified provided $\gamma \gg 1$. If say $\gamma = 10$ and $\delta$ is the same as for data displayed in Fig.~\ref{fig:contact_th}, \eqref{eq:CC} yields $ \ell _{GC} \simeq 2.65$ nm. However, using $\delta = 0.2$ nm, which is the smallest hydrodynamic diameter of inorganic ions, we obtain $ \ell _{GC} \simeq 1$ nm, which corresponds to very strongly charged surfaces.

\begin{figure}[h]
	\centering
	\includegraphics[width=0.99\columnwidth ]{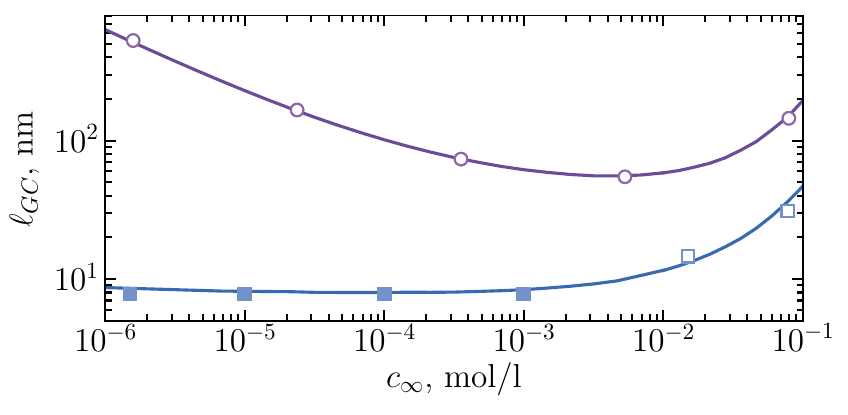}
	\caption{The Gouy-Chapman length $\ell_{GC}$ as a function of $c_{\infty}$. Solid curves from top to bottom are computed for $\phi_{0} = 1$ and $15$ using $\delta = 0.5$ nm and $\gamma = 40$. Open circles show predictions of Eq.~\eqref{eq:small_phi0}. Filled and open squares are calculations from Eqs.~\eqref{eq:CC} and \eqref{eq:charge_fin}. }
	\label{fig:lgc_salt}
\end{figure}

Probably the most striking conclusion from Eq.~\eqref{eq:CC} is that in this mode the surface charge density does not change with the variation of salt concentration. This implies that the surface of a conductor obeys the (Neumann) boundary condition of a constant charge that would normally be expected for charged insulators. This situation corresponds to the plateau of the Gouy-Chapman length in Fig.~\ref{fig:lgc_salt} observed for $\phi_0 = 15$ in dilute solutions ($c_{\infty} \leq 10^{-3}$ mol/l). However, in more concentrated solutions at high $\phi_0$ the Gouy-Chapman length grows exponentially with salt, and we return to this later.
Also included in Fig.~\ref{fig:lgc_salt} the Gouy-Chapman length calculated for $\phi_0 = 1$. In this case  $\ell_{GC}$ is much larger, the plateau disappears and has its minimum at $c_{\infty}\simeq 10^{-2}$ mol/l (where $\lambda \simeq 3$ nm and $\sqrt{\gamma}\delta/\lambda \simeq 1$), although an effective surface charge $\lambda/\ell_{GC}$ (not shown) and $\phi_s$ (see Fig.~\ref{fig:delta}) decrease monotonously with the bulk concentration. This result implies that the surface charge density plotted against $c_{\infty}$ always exhibits a  maximum, and that  surfaces bearing different potentials can be of the same $\sigma$. An explanation can be obtained if we invoke asymptotic limits of Eq.~\eqref{eq:small_phi0}.
From Eq.~\eqref{eq:uematsu} it follows that \eqref{eq:small_phi0} reduces to
\begin{equation}
  \ell_{GC} \simeq  \dfrac{2 \lambda(1 + \gamma \delta/\lambda)}{\phi_0}
\end{equation}
in very dilute solutions, which implies it decreases with salt as $ \ell_{GC} \propto 1/\sqrt{c_{\infty}}$. This is exactly the shape of the left (decaying) branch of the $\ell_{GC}$-curve. For the right-side branch $\sqrt{\gamma}\delta/\lambda \geq 1$, so that using Eqs.~\eqref{eq:phis_small_general2} and \eqref{eq:Grahame_smallphi} one can derive
\begin{equation}\label{eq:small_lambda}
   \ell_{GC} \simeq \dfrac{\lambda (1 + \sqrt{ \gamma})e^{\sqrt{ \gamma} \delta/\lambda}}{\phi_0 }.
  \end{equation}
which decreases with $\lambda$ (since it is smaller than $\sqrt{\gamma}\delta$), \emph{i.e.} augments with concentration as $\ell_{GC}\propto  e^{\sqrt{c_{\infty}}}/\sqrt{c_{\infty}}$. We recall that Eq.~\eqref{eq:small_lambda} refers to $\phi_0 \leq 1$. If $\phi_0 \geq 4$, but $\phi_s$ remains small, we move to the situation described by Eq.~\eqref{eq:lowphisat} and
\begin{equation}\label{eq:charge_fin}
   \ell_{GC} \simeq \dfrac{\lambda (1 + \sqrt{ \gamma})e^{\sqrt{ \gamma} \delta/\lambda}}{4}.
  \end{equation}
 Thus this case yields the same scaling of the Gouy-Chapman length with salt as that of Eq.~\eqref{eq:small_lambda}, but the magnitude of  $\ell_{GC}$ is much smaller. Finally, we note that Eq.~\eqref{eq:charge_fin} allows one to interpret the right-side branch ($c_{\infty}\geq 10^{-2}$ mol/l) of the lower curve in Fig.~\ref{fig:lgc_salt}.


\section{Final remarks}

We propose a mean-field theory describing the surface potential of conductors in electrolyte solutions that incorporates the effects of reduced permittivity of the Stern layer  coupled with the mobility of its ions.
Our model produces an endorsement of numerous
experimental observations including the control of the surface potential by varying the electrical potential applied to the conductor~\cite{raiteri.r:1998,connor.jn:2001}, an abrupt jump between the electrode and surface  potentials~\cite{frechette.j:2001,connor.jn:2002}, the saturation of the surface charge density at high applied voltage~\cite{lyklema:2010}, and more. Although these experimental results have long been known, a direct connection between the two potentials has not previously been formulated. Thus perhaps the most important aspect of our work are the simple analytic relations between the electrode  and surface potentials found for experimentally relevant situations.  They have been derived by using solely electrostatic arguments. The    need    to    invoke    specific    chemisorption  is thereby  removed.

Our model provides considerable insight into an electrostatic boundary condition at the conductor surface.  It is usually believed that at the surface of a conducting body the constant potential condition should be imposed, although electrodes and colloidal particles normally do not conform to this standard idea. Our results show that for conductors immersed in electrolyte solutions this classical condition never hold. Their surface potential is adaptive being regulated by salt concentration. This result has large implications  and represents a step forward. The same concerns the surface charge density of the electrode, which is also established self-consistently. The consequences are huge. For instance, in some situations the conductor might obey a condition of a fixed density of the surface charge normally expected for insulators, or say the same surface charge density may correspond to different surface potentials. In essence, the adaptive boundary conditions is a universal phenomenon not accommodated by standard theories yet. Thus our results may guide the re-interpreting the inner Stern layer capacity and electrokinetic phenomena near isolated walls or in thick channels, where EDLs do not overlap.

Our analysis was motivated in part by the surface force measurements of~\citet{connor.jn:2001,connor.jn:2002} between a mercury drop and mica, as well as measurements employing other polarisable electrodes~\cite{raiteri.r:1998,frechette.j:2001,collins.l:2018}, that have been used to infer the surface potential of conductors. Although available data are still rather scarce and cannot give unambiguous conclusions regarding impact of salt on $\phi_s$, they lend some supports to the picture that is presented here. All measurements have been unsuccessful for $c_{\infty}$ above $10^{-3}$ mol/l due to a jump of interacting bodies into contact. This implies that $\phi_s$ becomes low, which is exactly what we have predicted. It would be of much interest to perform such measurements in solutions of different concentrations to see how does $\phi_s$ varies with separations, and in principle it should be  feasible to measure even in concentrated solutions by monitoring the jumps (into contact).  Our results show that $\Delta \phi = \phi_0 -\phi_s$ increases with $\phi_0$, and that $\phi_s$ saturates at high $\phi_0$. The increase in $\Delta \phi$ is well supported by experiments of \citet{frechette.j:2001} with the gold electrode. These authors also found that the surface force saturates as applied potential is increased and speculated that the force becomes insensitive to variations in $\phi_s$ when the latter is large. An alternative explanation could be a saturation of $\phi_s$ predicted here. Taking the thickness of the Stern layer $\delta=0.2$ nm, $\gamma=10$, and the experimental value of $\lambda\simeq 10$ nm,  one can find from Eq.~\eqref{eq:plateau} that $\phi_s$ at saturation is about 6 (or 150 mV), which is rather close to the onset of the saturation effect in experiment~\cite{frechette.j:2001}. However, the authors of experiments with mercury ($\lambda \simeq 10$~nm) concluded  that  $\phi_s$ does not saturate, although increases more and more slowly as potential is applied~\cite{connor.jn:2001}. This does not contradict to our results since their $\phi_0 \leq 12$ and the saturation mode is likely not reached yet, but the same conclusion has later been made for $\phi_0 = 20$ and lower $c_{\infty}$ ($\lambda \simeq 30$~nm)~\cite{connor.jn:2002}. A possible explanation for this discrepancy is a procedure to infer the surface potential, which besets with difficulties. Assuming that $\phi_s$ differs by a constant amount from the potential applied to the interior of the mercury, these authors found that the shifted by this amount data coincide with the Poisson-Boltzmann calculations within the standard, \emph{i.e.} without the Stern layer, model. Since there is still no convincing theoretical support for an assumption of the constant and independent on concentration $\Delta \phi$, we suggest further analysis of these measurements that employ the double layer model proposed here. Our current (single ``wall'') theory should, of course, be generalised to the case of a confined geometry, and this could be a fruitful direction. In particular, one can even speculate that our model could be in some cases an alternative to a so-called charge regulation approach~\cite{chan.d:1975,behrens.sh:2001,markovich.t:2016}, which is widely applied to interpret changes in electrostatic boundary conditions with the body inter-distance.

The advantage of the approach lies in the fact that it can be extended in various ways by including other effects even for a simplest  single ``wall'' geometry. Some extensions appear natural and rather straightforward. For instance, we have neglected a specific chemisorption, but if necessary, this could easily be added to the model to improve  the  fit to experimental data. Some publications concluded that the rise in the capacitance of a mercury electrode
is probably due, at any rate partly, to adsorbed hydroxyl ions~\cite{mott.nf:1961} of hydrodynamic diameter (0.09~nm~\cite{kadhim.mj:2020}),  which is of at least from 2 to 6 times smaller  than that of inorganic ions.  It would be straightforward to include them to the model to see what kind of quantitative or perhaps even qualitative changes might be expected.
More challenging extensions could be, for example, the use of the so-called Jellium model of charge distribution at a metal interface~\cite{guidelli.r:2000} instead of representation of the metal interface by an ideal (sharp) plane, or accounting for a dependence of the inner permittivity on the field~\cite{conway.be:1999}.
Finally, we note that since the Poisson-Boltzmann equation does not take into account the finite size of the adsorbed ions, the ionic concentration
in the Stern layer can potentially reach unrealistically high values. A modification of the
Poisson-Boltzmann equation normally allows a remedy of this problem in the outer layer~\cite{borukhov.i:1997}. It would be of interest to address this issue for the diffuse inner layer too.

\begin{acknowledgments}

This work was supported by the Ministry of Science and Higher Education of the Russian Federation.
We are indebted to  G.~A.~Tsirlina for drawing attention to some references, as well as to J.~Frechette and R.~G.~Horn for clarifying the details of their experiments.
\end{acknowledgments}

\section*{DATA AVAILABILITY}

The data that support the findings of this study are available within the
article.

\section*{AUTHORS' CONTRIBUTIONS}

O.I.V. designed and supervised the project, developed the
theory, and wrote the manuscript. E.F.S. developed numerical codes, performed computations,
and prepared the figures. E.S.A. participated in calculations and verified the theory.

\section*{AUTHOR DECLARATIONS}

The authors have no conflicts to disclose.


\bibliography{stern}

\begin{thebibliography}{42}%
\makeatletter
\providecommand \@ifxundefined [1]{%
 \@ifx{#1\undefined}
}%
\providecommand \@ifnum [1]{%
 \ifnum #1\expandafter \@firstoftwo
 \else \expandafter \@secondoftwo
 \fi
}%
\providecommand \@ifx [1]{%
 \ifx #1\expandafter \@firstoftwo
 \else \expandafter \@secondoftwo
 \fi
}%
\providecommand \natexlab [1]{#1}%
\providecommand \enquote  [1]{``#1''}%
\providecommand \bibnamefont  [1]{#1}%
\providecommand \bibfnamefont [1]{#1}%
\providecommand \citenamefont [1]{#1}%
\providecommand \href@noop [0]{\@secondoftwo}%
\providecommand \href [0]{\begingroup \@sanitize@url \@href}%
\providecommand \@href[1]{\@@startlink{#1}\@@href}%
\providecommand \@@href[1]{\endgroup#1\@@endlink}%
\providecommand \@sanitize@url [0]{\catcode `\\12\catcode `\$12\catcode
  `\&12\catcode `\#12\catcode `\^12\catcode `\_12\catcode `\%12\relax}%
\providecommand \@@startlink[1]{}%
\providecommand \@@endlink[0]{}%
\providecommand \url  [0]{\begingroup\@sanitize@url \@url }%
\providecommand \@url [1]{\endgroup\@href {#1}{\urlprefix }}%
\providecommand \urlprefix  [0]{URL }%
\providecommand \Eprint [0]{\href }%
\providecommand \doibase [0]{https://doi.org/}%
\providecommand \selectlanguage [0]{\@gobble}%
\providecommand \bibinfo  [0]{\@secondoftwo}%
\providecommand \bibfield  [0]{\@secondoftwo}%
\providecommand \translation [1]{[#1]}%
\providecommand \BibitemOpen [0]{}%
\providecommand \bibitemStop [0]{}%
\providecommand \bibitemNoStop [0]{.\EOS\space}%
\providecommand \EOS [0]{\spacefactor3000\relax}%
\providecommand \BibitemShut  [1]{\csname bibitem#1\endcsname}%
\let\auto@bib@innerbib\@empty
\bibitem [{\citenamefont {Stern}(1924)}]{stern.o:1924}%
  \BibitemOpen
  \bibfield  {author} {\bibinfo {author} {\bibfnamefont {O.}~\bibnamefont
  {Stern}},\ }\bibfield  {title} {\bibinfo {title} {Zur {T}heorie der
  elektrolytischen {D}oppelschicht},\ }\href@noop {} {\bibfield  {journal}
  {\bibinfo  {journal} {Z. Elektrochem}\ }\textbf {\bibinfo {volume} {30}},\
  \bibinfo {pages} {1014} (\bibinfo {year} {1924})}\BibitemShut {NoStop}%
\bibitem [{\citenamefont {Connor}\ and\ \citenamefont
  {Horn}(2001)}]{connor.jn:2001}%
  \BibitemOpen
  \bibfield  {author} {\bibinfo {author} {\bibfnamefont {J.~N.}\ \bibnamefont
  {Connor}}\ and\ \bibinfo {author} {\bibfnamefont {R.~G.}\ \bibnamefont
  {Horn}},\ }\bibfield  {title} {\bibinfo {title} {Measurement of aqueous film
  thickness between charged mercury and mica surfaces: {{A}} direct
  experimental probe of the {P}oisson--{B}oltzmann distribution},\ }\href@noop
  {} {\bibfield  {journal} {\bibinfo  {journal} {Langmuir}\ }\textbf {\bibinfo
  {volume} {17}},\ \bibinfo {pages} {7194} (\bibinfo {year}
  {2001})}\BibitemShut {NoStop}%
\bibitem [{\citenamefont {Mott}\ and\ \citenamefont
  {Watts-Tobin}(1961)}]{mott.nf:1961}%
  \BibitemOpen
  \bibfield  {author} {\bibinfo {author} {\bibfnamefont {N.}~\bibnamefont
  {Mott}}\ and\ \bibinfo {author} {\bibfnamefont {R.}~\bibnamefont
  {Watts-Tobin}},\ }\bibfield  {title} {\bibinfo {title} {The interface between
  a metal and an electrolyte},\ }\href@noop {} {\bibfield  {journal} {\bibinfo
  {journal} {Electrochimica Acta}\ }\textbf {\bibinfo {volume} {4}},\ \bibinfo
  {pages} {79} (\bibinfo {year} {1961})}\BibitemShut {NoStop}%
\bibitem [{\citenamefont {Conway}(1999)}]{conway.be:1999}%
  \BibitemOpen
  \bibfield  {author} {\bibinfo {author} {\bibfnamefont {B.~E.}\ \bibnamefont
  {Conway}},\ }\bibinfo {title} {Theoretical treatment and modeling of the
  double layer at electrode interfaces},\ in\ \href
  {https://doi.org/10.1007/978-1-4757-3058-6_7} {\emph {\bibinfo {booktitle}
  {Electrochemical Supercapacitors: Scientific Fundamentals and Technological
  Applications}}}\ (\bibinfo  {publisher} {Springer US},\ \bibinfo {address}
  {Boston, MA},\ \bibinfo {year} {1999})\ pp.\ \bibinfo {pages}
  {125--168}\BibitemShut {NoStop}%
\bibitem [{\citenamefont {Biesheuvel}\ and\ \citenamefont
  {Dykstra}(2021)}]{biesheuvel.pm:2021}%
  \BibitemOpen
  \bibfield  {author} {\bibinfo {author} {\bibfnamefont {P.~M.}\ \bibnamefont
  {Biesheuvel}}\ and\ \bibinfo {author} {\bibfnamefont {J.~E.}\ \bibnamefont
  {Dykstra}},\ }\href@noop {} {\emph {\bibinfo {title} {Introduction to Physics
  of Electrochemical Processes}}}\ (\bibinfo  {publisher}
  {http://www.physicsofelectrochemicalprocesses.com ISBN: 9789090341064},\
  \bibinfo {year} {2021})\BibitemShut {NoStop}%
\bibitem [{\citenamefont {Derjaguin}\ and\ \citenamefont
  {Landau}(1941)}]{derjaguin.bv:1941}%
  \BibitemOpen
  \bibfield  {author} {\bibinfo {author} {\bibfnamefont {B.~V.}\ \bibnamefont
  {Derjaguin}}\ and\ \bibinfo {author} {\bibfnamefont {L.~D.}\ \bibnamefont
  {Landau}},\ }\bibfield  {title} {\bibinfo {title} {Theory of the stability of
  strongly charged lyophobic sols and of the adhesion of strongly charged
  particles in solutions of electrolytes},\ }\href@noop {} {\bibfield
  {journal} {\bibinfo  {journal} {Acta Physicochimica U.R.S.S.}\ }\textbf
  {\bibinfo {volume} {14}},\ \bibinfo {pages} {633} (\bibinfo {year}
  {1941})}\BibitemShut {NoStop}%
\bibitem [{\citenamefont {Vervey}\ and\ \citenamefont
  {Overbeek}(1948)}]{vervey:1948}%
  \BibitemOpen
  \bibfield  {author} {\bibinfo {author} {\bibfnamefont {E.~J.~W.}\
  \bibnamefont {Vervey}}\ and\ \bibinfo {author} {\bibfnamefont {J.~T.~G.}\
  \bibnamefont {Overbeek}},\ }\href@noop {} {\emph {\bibinfo {title} {Theory of
  the stability of lyophobic colloids. The interaction of particles having an
  electric double layer}}}\ (\bibinfo  {publisher} {Elsevier},\ \bibinfo
  {address} {New York -- Amsterdam},\ \bibinfo {year} {1948})\BibitemShut
  {NoStop}%
\bibitem [{\citenamefont {Israelachvili}(2011)}]{israelachvili.jn:2011}%
  \BibitemOpen
  \bibfield  {author} {\bibinfo {author} {\bibfnamefont {J.~N.}\ \bibnamefont
  {Israelachvili}},\ }\href@noop {} {\emph {\bibinfo {title} {Intermolecular
  and Surface Forces}}},\ \bibinfo {edition} {3rd}\ ed.\ (\bibinfo  {publisher}
  {Academic Press},\ \bibinfo {year} {2011})\BibitemShut {NoStop}%
\bibitem [{\citenamefont {Anderson}(1989)}]{anderson.jl:1989}%
  \BibitemOpen
  \bibfield  {author} {\bibinfo {author} {\bibfnamefont {J.~L.}\ \bibnamefont
  {Anderson}},\ }\bibfield  {title} {\bibinfo {title} {Colloid transport by
  interfacial forces},\ }\href@noop {} {\bibfield  {journal} {\bibinfo
  {journal} {Annu. Rev. Fluid Mech.}\ }\textbf {\bibinfo {volume} {21}},\
  \bibinfo {pages} {61} (\bibinfo {year} {1989})}\BibitemShut {NoStop}%
\bibitem [{\citenamefont {Nizkaya}\ \emph {et~al.}(2022)\citenamefont
  {Nizkaya}, \citenamefont {Asmolov},\ and\ \citenamefont
  {Vinogradova}}]{nizkaya.tv:2022}%
  \BibitemOpen
  \bibfield  {author} {\bibinfo {author} {\bibfnamefont {T.~V.}\ \bibnamefont
  {Nizkaya}}, \bibinfo {author} {\bibfnamefont {E.~S.}\ \bibnamefont
  {Asmolov}},\ and\ \bibinfo {author} {\bibfnamefont {O.~I.}\ \bibnamefont
  {Vinogradova}},\ }\bibfield  {title} {\bibinfo {title} {Theoretical modeling
  of catalytic self-propulsion},\ }\href@noop {} {\bibfield  {journal}
  {\bibinfo  {journal} {Curr. Opin. Colloid Interface Sci.}\ }\textbf {\bibinfo
  {volume} {62}},\ \bibinfo {pages} {101637} (\bibinfo {year}
  {2022})}\BibitemShut {NoStop}%
\bibitem [{\citenamefont {Vinogradova}(2024)}]{vinogradova.oi:2024}%
  \BibitemOpen
  \bibfield  {author} {\bibinfo {author} {\bibfnamefont {O.~I.}\ \bibnamefont
  {Vinogradova}},\ }\href@noop {} {\emph {\bibinfo {title} {Colloidal systems:
  colloid ``chemistry'' for physicists}}}\ (\bibinfo  {publisher} {Moscow
  University Publishing House, ISBN: 978-5-19-011962-6},\ \bibinfo {address}
  {Moscow},\ \bibinfo {year} {2024})\BibitemShut {NoStop}%
\bibitem [{\citenamefont {Andelman}(2006)}]{andelman.d:2006book}%
  \BibitemOpen
  \bibfield  {author} {\bibinfo {author} {\bibfnamefont {D.}~\bibnamefont
  {Andelman}},\ }\bibinfo {title} {Introduction to electrostatics in soft and
  biological matter},\ in\ \href@noop {} {\emph {\bibinfo {booktitle} {Soft
  Condensed Matter Physics in Molecular and Cell Biology}}},\ \bibinfo {editor}
  {edited by\ \bibinfo {editor} {\bibfnamefont {W.}~\bibnamefont {Poon}}\ and\
  \bibinfo {editor} {\bibfnamefont {D.}~\bibnamefont {Andelman}}}\ (\bibinfo
  {publisher} {CRC Press},\ \bibinfo {address} {Boca Raton},\ \bibinfo {year}
  {2006})\ Chap.~\bibinfo {chapter} {6}, pp.\ \bibinfo {pages}
  {97--122}\BibitemShut {NoStop}%
\bibitem [{\citenamefont {Herrero}\ and\ \citenamefont
  {Joly}(2021)}]{joly.l:PB}%
  \BibitemOpen
  \bibfield  {author} {\bibinfo {author} {\bibfnamefont {C.}~\bibnamefont
  {Herrero}}\ and\ \bibinfo {author} {\bibfnamefont {L.}~\bibnamefont {Joly}},\
  }\bibfield  {title} {\bibinfo {title} {Poisson-{B}oltzmann formulary},\
  }\href@noop {} {\bibfield  {journal} {\bibinfo  {journal} {arXiv preprint
  arXiv:2105.00720}\ } (\bibinfo {year} {2021})}\BibitemShut {NoStop}%
\bibitem [{\citenamefont {Bedzyk}\ \emph {et~al.}(1990)\citenamefont {Bedzyk},
  \citenamefont {Bommarito}, \citenamefont {Caffrey},\ and\ \citenamefont
  {Penner}}]{bedzyk.mj:1990}%
  \BibitemOpen
  \bibfield  {author} {\bibinfo {author} {\bibfnamefont {M.~J.}\ \bibnamefont
  {Bedzyk}}, \bibinfo {author} {\bibfnamefont {J.~M.}\ \bibnamefont
  {Bommarito}}, \bibinfo {author} {\bibfnamefont {M.}~\bibnamefont {Caffrey}},\
  and\ \bibinfo {author} {\bibfnamefont {T.~L.}\ \bibnamefont {Penner}},\
  }\bibfield  {title} {\bibinfo {title} {Diffuse-double layer at a
  membrane-aqueous interface measured with {X}-ray standing waves},\
  }\href@noop {} {\bibfield  {journal} {\bibinfo  {journal} {Science}\ }\textbf
  {\bibinfo {volume} {248}},\ \bibinfo {pages} {52} (\bibinfo {year}
  {1990})}\BibitemShut {NoStop}%
\bibitem [{\citenamefont {Hartkamp}\ \emph {et~al.}(2018)\citenamefont
  {Hartkamp}, \citenamefont {Biance}, \citenamefont {Fu}, \citenamefont
  {Dufreche}, \citenamefont {Bonhomme},\ and\ \citenamefont
  {Joly}}]{hartkamp.r:2018}%
  \BibitemOpen
  \bibfield  {author} {\bibinfo {author} {\bibfnamefont {R.}~\bibnamefont
  {Hartkamp}}, \bibinfo {author} {\bibfnamefont {A.-L.}\ \bibnamefont
  {Biance}}, \bibinfo {author} {\bibfnamefont {L.}~\bibnamefont {Fu}}, \bibinfo
  {author} {\bibfnamefont {J.-F.}\ \bibnamefont {Dufreche}}, \bibinfo {author}
  {\bibfnamefont {O.}~\bibnamefont {Bonhomme}},\ and\ \bibinfo {author}
  {\bibfnamefont {L.}~\bibnamefont {Joly}},\ }\bibfield  {title} {\bibinfo
  {title} {Measuring surface charge: why experimental characterization and
  molecular modeling should be coupled},\ }\href@noop {} {\bibfield  {journal}
  {\bibinfo  {journal} {Curr. Opin. Colloid Interface Sci.}\ } (\bibinfo {year}
  {2018})}\BibitemShut {NoStop}%
\bibitem [{\citenamefont {Lyklema}(2010)}]{lyklema:2010}%
  \BibitemOpen
  \bibfield  {author} {\bibinfo {author} {\bibfnamefont {J.}~\bibnamefont
  {Lyklema}},\ }\bibfield  {title} {\bibinfo {title} {Molecular interpretation
  of electrokinetic potentials},\ }\href@noop {} {\bibfield  {journal}
  {\bibinfo  {journal} {Curr. Opin. Colloid Interface Sci.}\ }\textbf {\bibinfo
  {volume} {15}},\ \bibinfo {pages} {125} (\bibinfo {year} {2010})}\BibitemShut
  {NoStop}%
\bibitem [{\citenamefont {Conway}\ \emph {et~al.}(1951)\citenamefont {Conway},
  \citenamefont {Bockris},\ and\ \citenamefont {Ammar}}]{conway.be:1951}%
  \BibitemOpen
  \bibfield  {author} {\bibinfo {author} {\bibfnamefont {B.~E.}\ \bibnamefont
  {Conway}}, \bibinfo {author} {\bibfnamefont {J.~O.}\ \bibnamefont
  {Bockris}},\ and\ \bibinfo {author} {\bibfnamefont {I.~A.}\ \bibnamefont
  {Ammar}},\ }\bibfield  {title} {\bibinfo {title} {The dielectric constant of
  the solution in the diffuse and {H}elmholtz double layers at a charged
  interface in aqueous solution},\ }\href@noop {} {\bibfield  {journal}
  {\bibinfo  {journal} {Trans. Faraday Soc.}\ }\textbf {\bibinfo {volume}
  {47}},\ \bibinfo {pages} {756} (\bibinfo {year} {1951})}\BibitemShut
  {NoStop}%
\bibitem [{\citenamefont {Borisevich}\ \emph {et~al.}(1999)\citenamefont
  {Borisevich}, \citenamefont {Kharkats},\ and\ \citenamefont
  {Tsirlina}}]{borisevich.sv:1999}%
  \BibitemOpen
  \bibfield  {author} {\bibinfo {author} {\bibfnamefont {S.~V.}\ \bibnamefont
  {Borisevich}}, \bibinfo {author} {\bibfnamefont {Y.~I.}\ \bibnamefont
  {Kharkats}},\ and\ \bibinfo {author} {\bibfnamefont {G.~A.}\ \bibnamefont
  {Tsirlina}},\ }\bibfield  {title} {\bibinfo {title} {Localization of the
  reagents at electrochemical interfaces and its role in the solvent
  reorganization},\ }\href@noop {} {\bibfield  {journal} {\bibinfo  {journal}
  {Russ. J. Electrochem.}\ }\textbf {\bibinfo {volume} {35}},\ \bibinfo {pages}
  {675} (\bibinfo {year} {1999})}\BibitemShut {NoStop}%
\bibitem [{\citenamefont {Guidelli}\ and\ \citenamefont
  {Schmickler}(2000)}]{guidelli.r:2000}%
  \BibitemOpen
  \bibfield  {author} {\bibinfo {author} {\bibfnamefont {R.}~\bibnamefont
  {Guidelli}}\ and\ \bibinfo {author} {\bibfnamefont {W.}~\bibnamefont
  {Schmickler}},\ }\bibfield  {title} {\bibinfo {title} {Recent developments in
  models for the interface between a metal and an aqueous solution},\
  }\href@noop {} {\bibfield  {journal} {\bibinfo  {journal} {Electrochimica
  Acta}\ }\textbf {\bibinfo {volume} {45}},\ \bibinfo {pages} {2317} (\bibinfo
  {year} {2000})}\BibitemShut {NoStop}%
\bibitem [{\citenamefont {Schlaich}\ \emph {et~al.}(2016)\citenamefont
  {Schlaich}, \citenamefont {Knapp},\ and\ \citenamefont
  {Netz}}]{schlaich.a:2016}%
  \BibitemOpen
  \bibfield  {author} {\bibinfo {author} {\bibfnamefont {A.}~\bibnamefont
  {Schlaich}}, \bibinfo {author} {\bibfnamefont {E.~W.}\ \bibnamefont
  {Knapp}},\ and\ \bibinfo {author} {\bibfnamefont {R.~R.}\ \bibnamefont
  {Netz}},\ }\bibfield  {title} {\bibinfo {title} {Water dielectric effects in
  planar confinement},\ }\href@noop {} {\bibfield  {journal} {\bibinfo
  {journal} {Phys. Rev. Lett.}\ }\textbf {\bibinfo {volume} {117}},\ \bibinfo
  {pages} {048001} (\bibinfo {year} {2016})}\BibitemShut {NoStop}%
\bibitem [{\citenamefont {Becker}\ \emph {et~al.}(2023)\citenamefont {Becker},
  \citenamefont {Loche}, \citenamefont {Rezaei}, \citenamefont {Wolde-Kidan},
  \citenamefont {Uematsu}, \citenamefont {Netz},\ and\ \citenamefont
  {Bonthuis}}]{becker.m:2023}%
  \BibitemOpen
  \bibfield  {author} {\bibinfo {author} {\bibfnamefont {M.}~\bibnamefont
  {Becker}}, \bibinfo {author} {\bibfnamefont {P.}~\bibnamefont {Loche}},
  \bibinfo {author} {\bibfnamefont {M.}~\bibnamefont {Rezaei}}, \bibinfo
  {author} {\bibfnamefont {A.}~\bibnamefont {Wolde-Kidan}}, \bibinfo {author}
  {\bibfnamefont {Y.}~\bibnamefont {Uematsu}}, \bibinfo {author} {\bibfnamefont
  {R.~R.}\ \bibnamefont {Netz}},\ and\ \bibinfo {author} {\bibfnamefont
  {D.~J.}\ \bibnamefont {Bonthuis}},\ }\bibfield  {title} {\bibinfo {title}
  {Multiscale modeling of aqueous electric double layers},\ }\href@noop {}
  {\bibfield  {journal} {\bibinfo  {journal} {Chem. Rev.}\ }\textbf {\bibinfo
  {volume} {124}},\ \bibinfo {pages} {1} (\bibinfo {year} {2023})}\BibitemShut
  {NoStop}%
\bibitem [{\citenamefont {Fumagalli}\ \emph {et~al.}(2018)\citenamefont
  {Fumagalli}, \citenamefont {Esfandiar}, \citenamefont {Fabregas},
  \citenamefont {Hu}, \citenamefont {Ares}, \citenamefont {Janardanan},
  \citenamefont {Yang}, \citenamefont {Radha}, \citenamefont {Taniguchi},
  \citenamefont {Watanabe} \emph {et~al.}}]{fumagalli.l:2018}%
  \BibitemOpen
  \bibfield  {author} {\bibinfo {author} {\bibfnamefont {L.}~\bibnamefont
  {Fumagalli}}, \bibinfo {author} {\bibfnamefont {A.}~\bibnamefont
  {Esfandiar}}, \bibinfo {author} {\bibfnamefont {R.}~\bibnamefont {Fabregas}},
  \bibinfo {author} {\bibfnamefont {S.}~\bibnamefont {Hu}}, \bibinfo {author}
  {\bibfnamefont {P.}~\bibnamefont {Ares}}, \bibinfo {author} {\bibfnamefont
  {A.}~\bibnamefont {Janardanan}}, \bibinfo {author} {\bibfnamefont
  {Q.}~\bibnamefont {Yang}}, \bibinfo {author} {\bibfnamefont {B.}~\bibnamefont
  {Radha}}, \bibinfo {author} {\bibfnamefont {T.}~\bibnamefont {Taniguchi}},
  \bibinfo {author} {\bibfnamefont {K.}~\bibnamefont {Watanabe}}, \emph
  {et~al.},\ }\bibfield  {title} {\bibinfo {title} {Anomalously low dielectric
  constant of confined water},\ }\href@noop {} {\bibfield  {journal} {\bibinfo
  {journal} {Science}\ }\textbf {\bibinfo {volume} {360}},\ \bibinfo {pages}
  {1339} (\bibinfo {year} {2018})}\BibitemShut {NoStop}%
\bibitem [{\citenamefont {Maduar}\ \emph {et~al.}(2015)\citenamefont {Maduar},
  \citenamefont {Belyaev}, \citenamefont {Lobaskin},\ and\ \citenamefont
  {Vinogradova}}]{maduar.sr:2015}%
  \BibitemOpen
  \bibfield  {author} {\bibinfo {author} {\bibfnamefont {S.~R.}\ \bibnamefont
  {Maduar}}, \bibinfo {author} {\bibfnamefont {A.~V.}\ \bibnamefont {Belyaev}},
  \bibinfo {author} {\bibfnamefont {V.}~\bibnamefont {Lobaskin}},\ and\
  \bibinfo {author} {\bibfnamefont {O.~I.}\ \bibnamefont {Vinogradova}},\
  }\bibfield  {title} {\bibinfo {title} {Electrohydrodynamics near hydrophobic
  surfaces},\ }\href@noop {} {\bibfield  {journal} {\bibinfo  {journal} {Phys.
  Rev. Lett.}\ }\textbf {\bibinfo {volume} {114}},\ \bibinfo {pages} {118301}
  (\bibinfo {year} {2015})}\BibitemShut {NoStop}%
\bibitem [{\citenamefont {Grosjean}\ \emph {et~al.}(2019)\citenamefont
  {Grosjean}, \citenamefont {Bocquet},\ and\ \citenamefont
  {Vuilleumier}}]{grosjean.b:2019}%
  \BibitemOpen
  \bibfield  {author} {\bibinfo {author} {\bibfnamefont {B.}~\bibnamefont
  {Grosjean}}, \bibinfo {author} {\bibfnamefont {M.-L.}\ \bibnamefont
  {Bocquet}},\ and\ \bibinfo {author} {\bibfnamefont {R.}~\bibnamefont
  {Vuilleumier}},\ }\bibfield  {title} {\bibinfo {title} {Versatile
  electrification of two-dimensional nanomaterials in water},\ }\href@noop {}
  {\bibfield  {journal} {\bibinfo  {journal} {Nat. Com.}\ }\textbf {\bibinfo
  {volume} {10}},\ \bibinfo {pages} {1656} (\bibinfo {year}
  {2019})}\BibitemShut {NoStop}%
\bibitem [{\citenamefont {Mouterde}\ \emph {et~al.}(2019)\citenamefont
  {Mouterde}, \citenamefont {Keerthi}, \citenamefont {Poggioli}, \citenamefont
  {Dar}, \citenamefont {Siria}, \citenamefont {Geim}, \citenamefont {Bocquet},\
  and\ \citenamefont {Radha}}]{mouterde.t:2019}%
  \BibitemOpen
  \bibfield  {author} {\bibinfo {author} {\bibfnamefont {T.}~\bibnamefont
  {Mouterde}}, \bibinfo {author} {\bibfnamefont {A.}~\bibnamefont {Keerthi}},
  \bibinfo {author} {\bibfnamefont {A.~R.}\ \bibnamefont {Poggioli}}, \bibinfo
  {author} {\bibfnamefont {S.~A.}\ \bibnamefont {Dar}}, \bibinfo {author}
  {\bibfnamefont {A.}~\bibnamefont {Siria}}, \bibinfo {author} {\bibfnamefont
  {A.~K.}\ \bibnamefont {Geim}}, \bibinfo {author} {\bibfnamefont
  {L.}~\bibnamefont {Bocquet}},\ and\ \bibinfo {author} {\bibfnamefont
  {B.}~\bibnamefont {Radha}},\ }\bibfield  {title} {\bibinfo {title} {Molecular
  streaming and its voltage control in angstr{\"o}m-scale channels},\
  }\href@noop {} {\bibfield  {journal} {\bibinfo  {journal} {Nature}\ }\textbf
  {\bibinfo {volume} {567}},\ \bibinfo {pages} {87} (\bibinfo {year}
  {2019})}\BibitemShut {NoStop}%
\bibitem [{\citenamefont {Mouterde}\ and\ \citenamefont
  {Bocquet}(2018)}]{mouterde.t:2018}%
  \BibitemOpen
  \bibfield  {author} {\bibinfo {author} {\bibfnamefont {T.}~\bibnamefont
  {Mouterde}}\ and\ \bibinfo {author} {\bibfnamefont {L.}~\bibnamefont
  {Bocquet}},\ }\bibfield  {title} {\bibinfo {title} {Interfacial transport
  with mobile surface charges and consequences for ionic transport in carbon
  nanotubes},\ }\href@noop {} {\bibfield  {journal} {\bibinfo  {journal} {Eur.
  Phys. J. E}\ }\textbf {\bibinfo {volume} {41}},\ \bibinfo {pages} {1}
  (\bibinfo {year} {2018})}\BibitemShut {NoStop}%
\bibitem [{\citenamefont {Silkina}\ \emph {et~al.}(2019)\citenamefont
  {Silkina}, \citenamefont {Asmolov},\ and\ \citenamefont
  {Vinogradova}}]{silkina.ef:2019}%
  \BibitemOpen
  \bibfield  {author} {\bibinfo {author} {\bibfnamefont {E.~F.}\ \bibnamefont
  {Silkina}}, \bibinfo {author} {\bibfnamefont {E.~S.}\ \bibnamefont
  {Asmolov}},\ and\ \bibinfo {author} {\bibfnamefont {O.~I.}\ \bibnamefont
  {Vinogradova}},\ }\bibfield  {title} {\bibinfo {title} {Electro-osmotic flow
  in hydrophobic nanochannels},\ }\href@noop {} {\bibfield  {journal} {\bibinfo
   {journal} {Phys. Chem. Chem. Phys.}\ }\textbf {\bibinfo {volume} {21}},\
  \bibinfo {pages} {23036} (\bibinfo {year} {2019})}\BibitemShut {NoStop}%
\bibitem [{\citenamefont {Vinogradova}\ \emph {et~al.}(2022)\citenamefont
  {Vinogradova}, \citenamefont {Silkina},\ and\ \citenamefont
  {Asmolov}}]{vinogradova.oi:2022}%
  \BibitemOpen
  \bibfield  {author} {\bibinfo {author} {\bibfnamefont {O.~I.}\ \bibnamefont
  {Vinogradova}}, \bibinfo {author} {\bibfnamefont {E.~F.}\ \bibnamefont
  {Silkina}},\ and\ \bibinfo {author} {\bibfnamefont {E.~S.}\ \bibnamefont
  {Asmolov}},\ }\bibfield  {title} {\bibinfo {title} {Transport of ions in
  hydrophobic nanotubes},\ }\href@noop {} {\bibfield  {journal} {\bibinfo
  {journal} {Phys. Fluids}\ }\textbf {\bibinfo {volume} {34}} (\bibinfo {year}
  {2022})}\BibitemShut {NoStop}%
\bibitem [{\citenamefont {Mangaud}\ \emph {et~al.}(2022)\citenamefont
  {Mangaud}, \citenamefont {Bocquet}, \citenamefont {Bocquet},\ and\
  \citenamefont {Rotenberg}}]{mangaud.e:2022}%
  \BibitemOpen
  \bibfield  {author} {\bibinfo {author} {\bibfnamefont {E.}~\bibnamefont
  {Mangaud}}, \bibinfo {author} {\bibfnamefont {M.-L.}\ \bibnamefont
  {Bocquet}}, \bibinfo {author} {\bibfnamefont {L.}~\bibnamefont {Bocquet}},\
  and\ \bibinfo {author} {\bibfnamefont {B.}~\bibnamefont {Rotenberg}},\
  }\bibfield  {title} {\bibinfo {title} {Chemisorbed vs physisorbed surface
  charge and its impact on electrokinetic transport: Carbon vs boron nitride
  surface},\ }\href@noop {} {\bibfield  {journal} {\bibinfo  {journal} {J.
  Chem. Phys.}\ }\textbf {\bibinfo {volume} {156}},\ \bibinfo {pages} {044703}
  (\bibinfo {year} {2022})}\BibitemShut {NoStop}%
\bibitem [{\citenamefont {Vinogradova}\ \emph {et~al.}(2023)\citenamefont
  {Vinogradova}, \citenamefont {Silkina},\ and\ \citenamefont
  {Asmolov}}]{vinogradova.oi:2023}%
  \BibitemOpen
  \bibfield  {author} {\bibinfo {author} {\bibfnamefont {O.~I.}\ \bibnamefont
  {Vinogradova}}, \bibinfo {author} {\bibfnamefont {E.~F.}\ \bibnamefont
  {Silkina}},\ and\ \bibinfo {author} {\bibfnamefont {E.~S.}\ \bibnamefont
  {Asmolov}},\ }\bibfield  {title} {\bibinfo {title} {Slippery and mobile
  hydrophobic electrokinetics: from single walls to nanochannels},\ }\href@noop
  {} {\bibfield  {journal} {\bibinfo  {journal} {Curr. Opin. Colloid Interface
  Sci.}\ }\textbf {\bibinfo {volume} {68}},\ \bibinfo {pages} {101742}
  (\bibinfo {year} {2023})}\BibitemShut {NoStop}%
\bibitem [{\citenamefont {Clasohm}\ \emph {et~al.}(2006)\citenamefont
  {Clasohm}, \citenamefont {Chen}, \citenamefont {Knoll}, \citenamefont
  {Vinogradova},\ and\ \citenamefont {Horn}}]{clasohm.ly:2006}%
  \BibitemOpen
  \bibfield  {author} {\bibinfo {author} {\bibfnamefont {L.~Y.}\ \bibnamefont
  {Clasohm}}, \bibinfo {author} {\bibfnamefont {M.}~\bibnamefont {Chen}},
  \bibinfo {author} {\bibfnamefont {W.}~\bibnamefont {Knoll}}, \bibinfo
  {author} {\bibfnamefont {O.~I.}\ \bibnamefont {Vinogradova}},\ and\ \bibinfo
  {author} {\bibfnamefont {R.~G.}\ \bibnamefont {Horn}},\ }\bibfield  {title}
  {\bibinfo {title} {Self-assembled monolayers on mercury probed in a modified
  surface force apparatus},\ }\href@noop {} {\bibfield  {journal} {\bibinfo
  {journal} {J. Phys. Chem. B}\ }\textbf {\bibinfo {volume} {110}},\ \bibinfo
  {pages} {25931} (\bibinfo {year} {2006})}\BibitemShut {NoStop}%
\bibitem [{\citenamefont {Uematsu}\ \emph {et~al.}(2018)\citenamefont
  {Uematsu}, \citenamefont {Netz},\ and\ \citenamefont
  {Bonthuis}}]{uematsu.y:2018}%
  \BibitemOpen
  \bibfield  {author} {\bibinfo {author} {\bibfnamefont {Y.}~\bibnamefont
  {Uematsu}}, \bibinfo {author} {\bibfnamefont {R.~R.}\ \bibnamefont {Netz}},\
  and\ \bibinfo {author} {\bibfnamefont {D.~J.}\ \bibnamefont {Bonthuis}},\
  }\bibfield  {title} {\bibinfo {title} {Analytical interfacial layer model for
  the capacitance and electrokinetics of charged aqueous interfaces},\
  }\href@noop {} {\bibfield  {journal} {\bibinfo  {journal} {Langmuir}\
  }\textbf {\bibinfo {volume} {34}},\ \bibinfo {pages} {9097} (\bibinfo {year}
  {2018})}\BibitemShut {NoStop}%
\bibitem [{\citenamefont {Kadhim}\ and\ \citenamefont
  {Gamaj}(2020)}]{kadhim.mj:2020}%
  \BibitemOpen
  \bibfield  {author} {\bibinfo {author} {\bibfnamefont {M.~J.}\ \bibnamefont
  {Kadhim}}\ and\ \bibinfo {author} {\bibfnamefont {M.~I.}\ \bibnamefont
  {Gamaj}},\ }\bibfield  {title} {\bibinfo {title} {Estimation of the diffusion
  coefficient and hydrodynamic radius ({S}tokes radius) for inorganic ions in
  solution depending on molar conductivity as electro-analytical technique-a
  review},\ }\href@noop {} {\bibfield  {journal} {\bibinfo  {journal} {J. Chem.
  Rev}\ }\textbf {\bibinfo {volume} {2}},\ \bibinfo {pages} {182} (\bibinfo
  {year} {2020})}\BibitemShut {NoStop}%
\bibitem [{\citenamefont {Butt}\ \emph {et~al.}(2003)\citenamefont {Butt},
  \citenamefont {Graf},\ and\ \citenamefont {Kappl}}]{butt.hj:2003}%
  \BibitemOpen
  \bibfield  {author} {\bibinfo {author} {\bibfnamefont {H.~J.}\ \bibnamefont
  {Butt}}, \bibinfo {author} {\bibfnamefont {K.}~\bibnamefont {Graf}},\ and\
  \bibinfo {author} {\bibfnamefont {M.}~\bibnamefont {Kappl}},\ }\href@noop {}
  {\emph {\bibinfo {title} {Physics and Chemistry of Interfaces}}}\ (\bibinfo
  {publisher} {Wiley‐VCH Verlag GmbH},\ \bibinfo {year} {2003})\BibitemShut
  {NoStop}%
\bibitem [{\citenamefont {Raiteri}\ \emph {et~al.}(1998)\citenamefont
  {Raiteri}, \citenamefont {Preuss}, \citenamefont {Grattarola},\ and\
  \citenamefont {Butt}}]{raiteri.r:1998}%
  \BibitemOpen
  \bibfield  {author} {\bibinfo {author} {\bibfnamefont {R.}~\bibnamefont
  {Raiteri}}, \bibinfo {author} {\bibfnamefont {M.}~\bibnamefont {Preuss}},
  \bibinfo {author} {\bibfnamefont {M.}~\bibnamefont {Grattarola}},\ and\
  \bibinfo {author} {\bibfnamefont {H.-J.}\ \bibnamefont {Butt}},\ }\bibfield
  {title} {\bibinfo {title} {Preliminary results on the electrostatic
  double-layer force between two surfaces with high surface potentials},\
  }\href@noop {} {\bibfield  {journal} {\bibinfo  {journal} {Colloids Surf. A:
  Physicochem. Eng. Asp}\ }\textbf {\bibinfo {volume} {136}},\ \bibinfo {pages}
  {191} (\bibinfo {year} {1998})}\BibitemShut {NoStop}%
\bibitem [{\citenamefont {Frechette}\ and\ \citenamefont
  {Vanderlick}(2001)}]{frechette.j:2001}%
  \BibitemOpen
  \bibfield  {author} {\bibinfo {author} {\bibfnamefont {J.}~\bibnamefont
  {Frechette}}\ and\ \bibinfo {author} {\bibfnamefont {T.~K.}\ \bibnamefont
  {Vanderlick}},\ }\bibfield  {title} {\bibinfo {title} {Double layer forces
  over large potential ranges as measured in an electrochemical surface forces
  apparatus},\ }\href@noop {} {\bibfield  {journal} {\bibinfo  {journal}
  {Langmuir}\ }\textbf {\bibinfo {volume} {17}},\ \bibinfo {pages} {7620}
  (\bibinfo {year} {2001})}\BibitemShut {NoStop}%
\bibitem [{\citenamefont {Connor}\ and\ \citenamefont
  {Horn}(2003)}]{connor.jn:2002}%
  \BibitemOpen
  \bibfield  {author} {\bibinfo {author} {\bibfnamefont {J.~N.}\ \bibnamefont
  {Connor}}\ and\ \bibinfo {author} {\bibfnamefont {R.~G.}\ \bibnamefont
  {Horn}},\ }\bibfield  {title} {\bibinfo {title} {The influence of surface
  forces on thin film drainage between a fluid drop and a flat solid},\
  }\href@noop {} {\bibfield  {journal} {\bibinfo  {journal} {Faraday Discuss.}\
  }\textbf {\bibinfo {volume} {123}},\ \bibinfo {pages} {193} (\bibinfo {year}
  {2003})}\BibitemShut {NoStop}%
\bibitem [{\citenamefont {Collins}\ \emph {et~al.}(2018)\citenamefont
  {Collins}, \citenamefont {Weber},\ and\ \citenamefont
  {Rodriguez}}]{collins.l:2018}%
  \BibitemOpen
  \bibfield  {author} {\bibinfo {author} {\bibfnamefont {L.}~\bibnamefont
  {Collins}}, \bibinfo {author} {\bibfnamefont {S.~A.}\ \bibnamefont {Weber}},\
  and\ \bibinfo {author} {\bibfnamefont {B.~J.}\ \bibnamefont {Rodriguez}},\
  }\bibinfo {title} {Applications of {KPFM}-based approaches for surface
  potential and electrochemical measurements in liquid},\ in\ \href@noop {}
  {\emph {\bibinfo {booktitle} {Kelvin Probe Force Microscopy: From Single
  Charge Detection to Device Characterization}}},\ \bibinfo {editor} {edited
  by\ \bibinfo {editor} {\bibfnamefont {S.}~\bibnamefont {Sadewasser}}\ and\
  \bibinfo {editor} {\bibfnamefont {T.}~\bibnamefont {Glatzel}}}\ (\bibinfo
  {publisher} {Springer International Publishing},\ \bibinfo {address} {Cham},\
  \bibinfo {year} {2018})\ pp.\ \bibinfo {pages} {391--433}\BibitemShut
  {NoStop}%
\bibitem [{\citenamefont {Chan}\ \emph {et~al.}(1975)\citenamefont {Chan},
  \citenamefont {Perram}, \citenamefont {White},\ and\ \citenamefont
  {Healy}}]{chan.d:1975}%
  \BibitemOpen
  \bibfield  {author} {\bibinfo {author} {\bibfnamefont {D.}~\bibnamefont
  {Chan}}, \bibinfo {author} {\bibfnamefont {J.}~\bibnamefont {Perram}},
  \bibinfo {author} {\bibfnamefont {L.}~\bibnamefont {White}},\ and\ \bibinfo
  {author} {\bibfnamefont {T.}~\bibnamefont {Healy}},\ }\bibfield  {title}
  {\bibinfo {title} {Regulation of surface potential at amphoteric surfaces
  during particle--particle interaction},\ }\href@noop {} {\bibfield  {journal}
  {\bibinfo  {journal} {J. Chem. Soc.{,} Faraday Trans. 1}\ }\textbf {\bibinfo
  {volume} {71}},\ \bibinfo {pages} {1046} (\bibinfo {year}
  {1975})}\BibitemShut {NoStop}%
\bibitem [{\citenamefont {Behrens}\ and\ \citenamefont
  {Grier}(2001)}]{behrens.sh:2001}%
  \BibitemOpen
  \bibfield  {author} {\bibinfo {author} {\bibfnamefont {S.~H.}\ \bibnamefont
  {Behrens}}\ and\ \bibinfo {author} {\bibfnamefont {D.~G.}\ \bibnamefont
  {Grier}},\ }\bibfield  {title} {\bibinfo {title} {The charge of glass and
  silica surfaces},\ }\href@noop {} {\bibfield  {journal} {\bibinfo  {journal}
  {J. Chem. Phys.}\ }\textbf {\bibinfo {volume} {115}},\ \bibinfo {pages}
  {6716} (\bibinfo {year} {2001})}\BibitemShut {NoStop}%
\bibitem [{\citenamefont {Markovich}\ \emph {et~al.}(2016)\citenamefont
  {Markovich}, \citenamefont {Andelman},\ and\ \citenamefont
  {Podgornik}}]{markovich.t:2016}%
  \BibitemOpen
  \bibfield  {author} {\bibinfo {author} {\bibfnamefont {T.}~\bibnamefont
  {Markovich}}, \bibinfo {author} {\bibfnamefont {D.}~\bibnamefont
  {Andelman}},\ and\ \bibinfo {author} {\bibfnamefont {R.}~\bibnamefont
  {Podgornik}},\ }\bibfield  {title} {\bibinfo {title} {Charge regulation: A
  generalized boundary condition?},\ }\href@noop {} {\bibfield  {journal}
  {\bibinfo  {journal} {EPL}\ }\textbf {\bibinfo {volume} {113}},\ \bibinfo
  {pages} {26004} (\bibinfo {year} {2016})}\BibitemShut {NoStop}%
\bibitem [{\citenamefont {Borukhov}\ \emph {et~al.}(1997)\citenamefont
  {Borukhov}, \citenamefont {Andelman},\ and\ \citenamefont
  {Orland}}]{borukhov.i:1997}%
  \BibitemOpen
  \bibfield  {author} {\bibinfo {author} {\bibfnamefont {I.}~\bibnamefont
  {Borukhov}}, \bibinfo {author} {\bibfnamefont {D.}~\bibnamefont {Andelman}},\
  and\ \bibinfo {author} {\bibfnamefont {H.}~\bibnamefont {Orland}},\
  }\bibfield  {title} {\bibinfo {title} {Steric effects in electrolytes: A
  modified {P}oisson-{B}oltzmann equation},\ }\href@noop {} {\bibfield
  {journal} {\bibinfo  {journal} {Phys. Rev. Lett.}\ }\textbf {\bibinfo
  {volume} {79}},\ \bibinfo {pages} {435} (\bibinfo {year} {1997})}\BibitemShut
  {NoStop}%
\end{thebibliography}%

\end{document}